\documentclass[prd,superscriptaddress,a4paper,showpacs,showkeys,nofootinbib]{revtex4}
\usepackage{graphicx}%  Default Latex 2eps package for embedding figures
\usepackage{subfigure}
\usepackage{rotating}
\usepackage{epsfig}
\usepackage[lofdepth,lotdepth,caption=false]{subfig}
\usepackage{bm}
\usepackage{amsmath}
\usepackage{dcolumn}% Align table columns on decimal point
\usepackage{bm}% bold math

\usepackage{url}
\usepackage{amsmath,amssymb,graphicx}
\usepackage{amsmath}
\usepackage{amssymb}
\usepackage{mathrsfs}
\usepackage{accents}
\usepackage{epsfig}
\usepackage{enumerate}

\usepackage{mathtools}
\usepackage{mathrsfs}

\usepackage{color}

\begin{document}

\begin{flushright}
MS-TP-22-25
\end{flushright}

\title{Implications of the QCD dynamics and a Super - Glashow astrophysical neutrino flux on  the description of ultrahigh energy neutrino data}

%\date{\today} 

\author{Victor P. {\sc Gon\c{c}alves}}
\email{barros@ufpel.edu.br}
\affiliation{Institut f\"ur Theoretische Physik, Westf\"alische Wilhelms-Universit\"at M\"unster,
Wilhelm-Klemm-Straße 9, D-48149 M\"unster, Germany
}
\affiliation{Institute of Modern Physics, Chinese Academy of Sciences,
  Lanzhou 730000, China}
\affiliation{Institute of Physics and Mathematics, Federal University of Pelotas, \\
  Postal Code 354,  96010-900, Pelotas, RS, Brazil}

\author{Diego R. {\sc Gratieri}}
\email{drgratieri@id.uff.br}
\affiliation{Escola de Engenharia Industrial Metal\'urgica de Volta Redonda,
Universidade Federal Fluminense (UFF),\\
 CEP 27255-125, Volta Redonda, RJ, Brazil}

\author{Alex S. C.  {\sc Quadros}}
\email{alexscq@gmail.com}
\affiliation{Institute of Physics and Mathematics, Federal University of Pelotas, \\
  Postal Code 354,  96010-900, Pelotas, RS, Brazil}


\begin{abstract}
The number of events observed in neutrino telescopes depends on the neutrino fluxes in the Earth, their absorption while crossing the Earth and their interaction in the detector. In this paper, we investigate the impact of the QCD dynamics at high energies on the energy dependence of the average inelasticity and angular dependence of the absorption probability during the neutrino propagation through the Earth, as well in the determination of the properties of the incident astrophysical  neutrino flux. Moreover, the number of events at the IceCube and IceCube - Gen2 are estimated considering different scenarios for the QCD dynamics and assuming the presence of a hypothetical Super - Glashow flux, which peaks for energies above the Glashow resonance.  
\end{abstract}
\maketitle

\section{Introduction}

The study of the ultrahigh-energy (UHE)  events in neutrino telescopes is expected to improve our understanding about the origin, propagation, and interaction of  neutrinos (for recent reviews see, e.g., Refs. \cite{Ackermann:2022rqc,Abraham:2022jse}). In recent years, the IceCube data has been used to constrain the energy behavior of the  astrophysical neutrino flux as well as to constrain the neutrino - hadron cross-section (See, e.g. Refs. \cite{
IceCube:2017roe,Bustamante:2017xuy,IceCube:2020rnc,Valera:2022ylt,Esteban:2022uuw}). Such studies are strongly motivated by the impact of these quantities on the event rate at the IceCube detector. UHE neutrino detectors do not directly measure neutrinos, but rather only secondary or even tertiary products of neutrino - induced showers  \cite{Ahlers:2018fkn}. The associated events can be typically classified into four topologies: through-going tracks,  cascades, starting tracks, and double cascades. Through - going tracks are created whenever a muon, some of them induced by a $\nu_{\mu}$ charged current (CC) interaction outside the detector, passes through the instrumented volume.  Cascade events arise from neutrino - induced particle electromagnetic and hadronic showers. Neutral - current (NC) interactions of any flavour generate hadronic showers. In contrast, in $\nu_e$ and $\nu_{\tau}$ CC interactions, the combined electromagnetic and hadronic showers are produced. Glashow resonance events produces a hadronic or an electromagnetic shower, depending on the decay channel.  When the muon is created by a $\nu_{\mu}$ CC interaction inside the instrumented volume, such event is denoted as a starting track. In this case, the topology will be characterized by a cascade of hadrons and a muon. Finally, for $\nu_{\tau}$ CC interactions, the tau produced will usually decay to an electron or hadrons, resulting in a second cascade displaced from the accompanying hadronic cascade. If these two cascades are observed inside the instrumented volume, the events are denoted "double bang". The characteristics of all these neutrino - induced showers are strongly dependent on  the inelasticity $Y$, which is the fraction of the neutrino energy transferred to the hadronic target in the laboratory frame. In particular, the inelasticity defines the relative sizes of the leptonic and hadronic showers induced in a charged current neutrino interaction and, as a consequence, its precise determination is fundamental  to extract  from the detected muon tracks or electromagnetic and hadronic showers, the  energy of the incident neutrino  in high energy neutrino telescopes. Another important aspect is that the energy behaviour of $Y$ is directly associated with the description of the neutrino - nucleon cross-section ($\sigma_{\nu N}$), which is expected to be modified at high energies by nonlinear corrections to the QCD dynamics~\cite{hdqcd} as well as to be sensitive to the presence of beyond Standard Model (BSM) Physics (For a recent study see, e.g. Ref.~\cite{Huang:2021mki}). 
Such dependence  is one of the motivations of the analysis performed in Ref. \cite{IceCube:2018pgc}, which have studied starting track events,  estimating the hadronic cascade and muon energies separately,  and measured the inelasticity distribution. The results obtained in Ref.  \cite{IceCube:2018pgc} indicate that the energy dependence of $Y$ in the energy range from $\approx 1$ to $\approx 100$ TeV is consistent with the SM predictions derived in Ref. \cite{CS} using the linear DGLAP evolution equations \cite{dglap}. An important open question is the behaviour of inelasticity in the energy range that will be studied in the next generation of neutrino telescopes and if the presence of new effects could be probed. One of our goals is to investigate the impact of nonlinear effects in the inelasticity at the energy range that will be probed by the  IceCube Gen2 \cite{IceCube-Gen2:2020qha}. In particular, we will consider the model proposed in Refs. \cite{Berger:2007ic,Block:2010ud,Block:2013mia,Block:2013nia}, denoted BBMT hereafter, which is based on the assumption that the proton structure function saturates the Froissart bound at high energies. Such approach takes into account the unitarity corrections at all orders in the strong hadronic interactions and provides a lower bound for $\sigma_{\nu N}$. The associated results will be compared with those derived assuming the validity of the linear DGLAP evolutions, as usually assumed in the theoretical and experimental studies.

Moreover, it is important to notice that the signals observed in UHE neutrino detectors  are also strongly dependent  on the neutrino fluxes incident at the Earth and their absorption during the passage through Earth to the detector (see, e.g., Refs. \cite{Vincent:2017svp,Jeong:2017mzv,Alvarez-Muniz:2018owm,Donini:2018tsg,Garcia:2020jwr}). One has that the attenuation of the incident neutrino flux depends on the neutrino energy and the arrival direction, with the neutrino propagation depending on the details of the matter structure between the source and the detector. For relatively small values of the neutrino energy ($E_{\nu} \lesssim 50$ TeV), the Earth is essentially transparent to neutrinos, while above it, the neutrinos traveling through a sufficient chord length inside the Earth may interact before  arriving at the detector.  
The description of this absorption is strongly dependent on $\sigma_{\nu N}$ and $Z(\theta_z)$, which is the total amount of matter that neutrino feel as a function of zenith angle $\theta_z$. Another goal of this paper is to study the impact of the nonlinear effects on the energy and angular dependencies of the probability $P_{Shad}(E_{\nu},\theta_{z})$ of neutrino interaction while crossing the Earth. In our analysis, we will compare the BBMT predictions with those derived using the linear DGLAP dynamics. For completeness, the predictions for $P_{Shad}(E_{\nu},\theta_{z})$ associated with the Glashow resonance will also be presented.

The  spectrum of astrophysical neutrinos is still being understood, with the predictions being greatly model dependent \cite{Coleman:2022abf}. A standard assumption is that it can be described by a single power law spectrum for all flavours, 
$\Phi_{astro} (E_{\nu}) = \Phi_{0}\times  \left( {E_{\nu}}/{E_{0}}   \right)^{-\gamma_{astro}}$, where  $\Phi_{0}$ is the flux normalization and $\gamma_{astro}$ is the spectral index \cite{IceCube:2018pgc,IceCube:2017zho,IceCube:2020wum,IceCube:2021uhz,IceCube:2020acn}. Considering the large theoretical uncertainty, it is common to assume $\Phi_{0}$ and $\gamma_{astro}$ as nuisance parameters in the analyses. %In particular, the study performed in Ref. \cite{IceCube:2018pgc} has obtained that   the energy spectrum of astrophysical neutrinos for both track and cascade samples can be described well by a power law with $\gamma_{astro} = 2.62 \pm 0.07$  in the energy range from 3.5 TeV to 2.6 PeV. {\color{red} , in tension with previous results based on through - going muons. SUBSTITUIR A PARTE EM VERMELHO POR::?} {\bf Indeed,
Recent results, derived in Ref. \cite{IceCube:2020acn}, favor a high energy flux with $\gamma_{astro}=2.3 - 2.6$, depending on the specific analysis. It is also important to emphasize that the IceCube Collaboration has also observed a Glashow resonance event which is consistent with this flux \cite{IceCube:2021rpz}. In order, to analyze the dependence of the astrophysical neutrino flux parameters on the description of the QCD dynamics and the treatment of the inelasticity, in this paper we also will perform a likelihood analysis of the number of neutrino events in six years of exposition of the HESE data \citep{Aartsen:2015knd} considering the distinct approaches for the calculation of the inelasticity described above. A comparison with the results obtained by IceCube will be presented.

Additionally, in recent years, some authors have discussed the possibility of a new astrophysical neutrino flux beyond the Glashow resonance, denoted Super-Glashow flux, which can generate a measurable number of ultraenergetic events (See, e.g. Refs. \cite{Kistler:2013my,Kistler:2016ask}). Such possibility is still hypothetical and theme of intense debate, mainly due to the fact the current data are quite well described disregarding a new component. However, as the magnitude of the nonlinear effects on the QCD dynamics increases with the energy, the presence of this new component in the astrophysical flux is expected to enhance the importance of a precise treatment of the neutrino - hadron interaction. In our analysis, we will assume the presence of this new hypothetical flux,  as proposed in Ref. \cite{Kistler:2016ask},  and  estimate the impact of these effects on the predictions for the number of events observed at the IceCube and IceCube - Gen2. For completeness, we will consider  five distinct combinations of parameters in the description of the Super-Glashow flux.

 This paper is organized as follows. In the next Section, we will present a brief review of the formalism needed to describe the average inelasticity in a neutrino - hadron interaction, the neutrino propagation in matter, and the calculation of the number of events observed in neutrino telescopes. Moreover, we will discuss  the  two models for $\sigma_{\nu N}$ assumed in our study. In Section \ref{Sec:res} we will present our predictions for the energy dependence of the average inelasticity considering the energy ranges covered by the IceCube and IceCube - Gen2. Moreover, we will present our results for the probability of  neutrino absorption by the Earth considering the distinct models for $\sigma_{\nu  N}$ and the different channels for the antineutrino - electron interaction. The impact on the properties of the astrophysical neutrino flux will be discussed. In addition, we will  analyze how the predictions for the number of events at the IceCube and IceCube - Gen2 are modified by the presence of a Super-Glashow flux. Finally, in Section \ref{sec:conc} we will summarize our main results and conclusions.

\begin{figure}[t]
\begin{center}
\begin{tabular}{cc}
\includegraphics[scale=0.36]{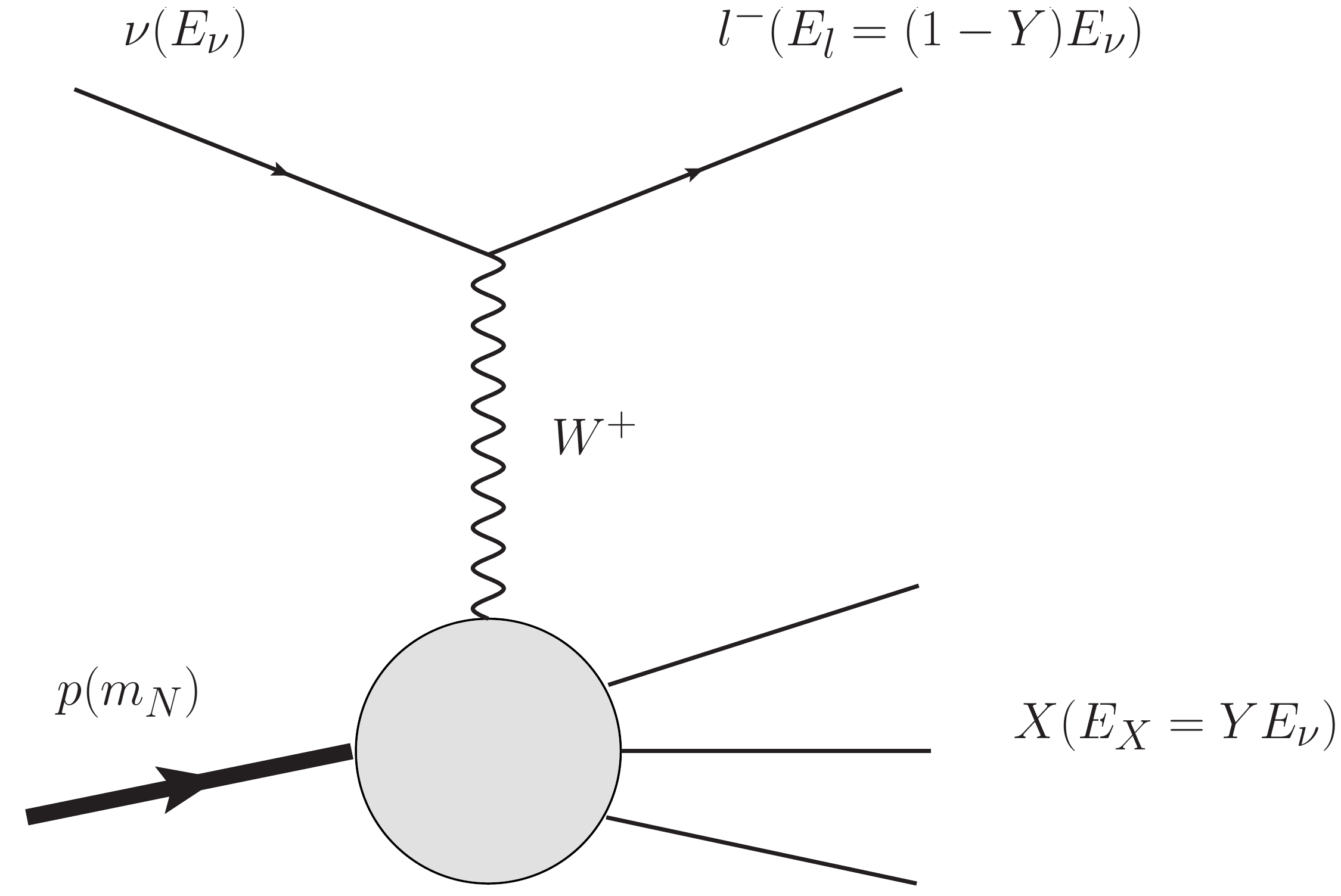} &  \includegraphics[scale=0.36]{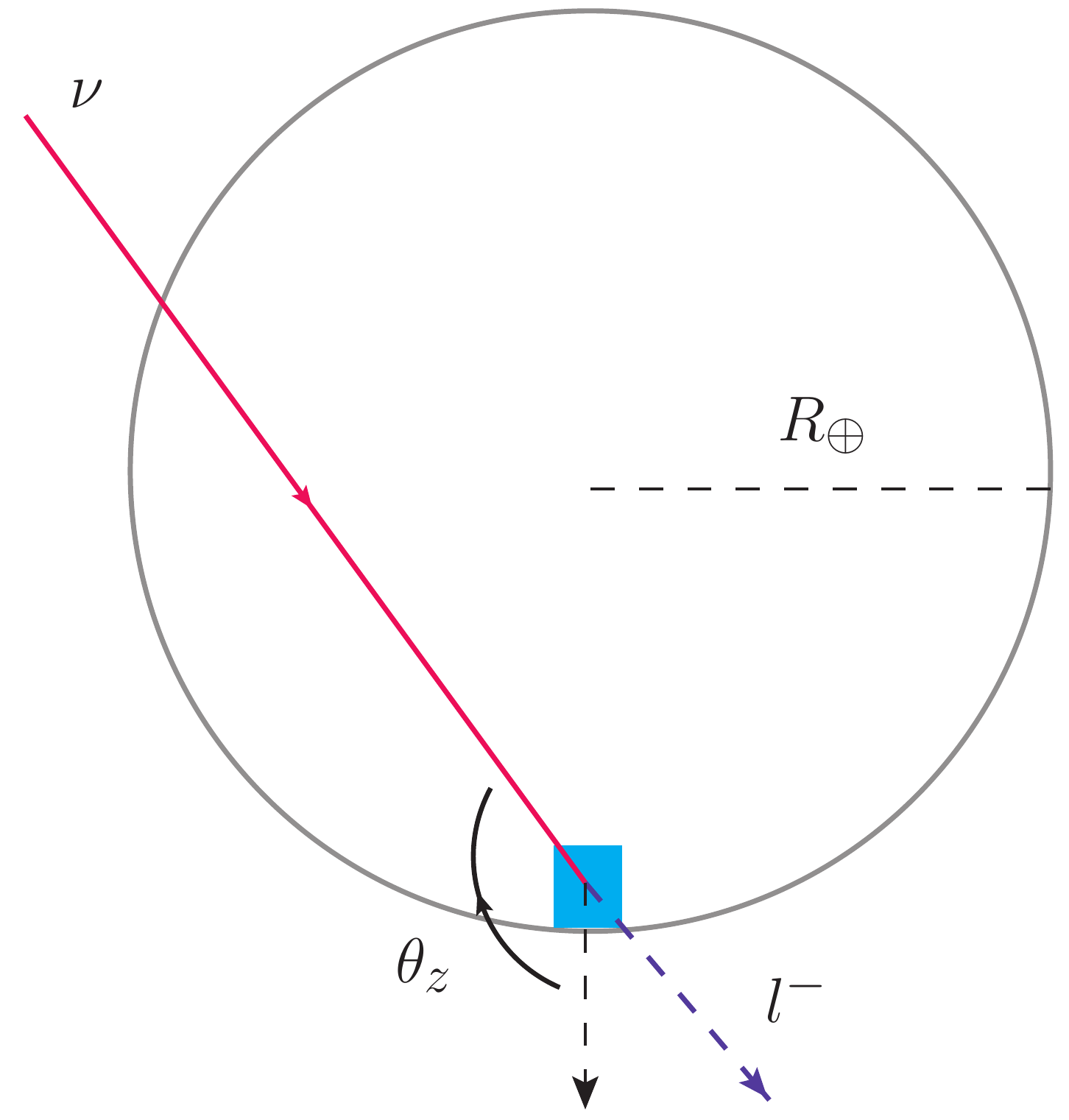} \\
(a) & (b)\\
\end{tabular}
\caption{(a) Deep inelastic neutrino - hadron scattering mediated by a $W$ exchange in the target rest frame; (b) Representation of the neutrino propagation through the Earth.}
\label{Fig:DIS}
\end{center}
\end{figure}

\section{Formalism}
\label{sec:form}

At ultrahigh energies, neutrinos interact mainly via deep inelastic scattering (DIS) \cite{book}. In this process,  the neutrino scatters off a quark in the nucleon via a virtual $W$ and $Z$ boson, producing a lepton and a hadronic system in the final state. The lepton present in the final state depends if one has a charged or a neutral current interaction. When the interaction is mediated by the $W$ boson, one has a {Charged Current}  (CC)   $\nu$-DIS,  and a charged lepton $l' = e, ~\mu,~ \tau$, is produced. On the other hand, if a $Z^0$ boson is exchanged and one has a { Neutral Current}  (NC) interaction, a neutrino of the same flavour than the incoming one will be present in the final state. 
The DIS processes can be completely described in terms of the four - momentum transfer $Q^2 \equiv - q^2$, where $q$ is the four - momentum of the gauge boson, the Bjorken - $x$ variable and inelasticity of the collision
 $Y$, which is  given in the target rest frame by 
$Y = E_X/E_{\nu}$, i.e. by the fraction of the neutrino's energy transferred to hadrons. As represented in Fig.  
 \ref{Fig:DIS}(a), the energies carried by the lepton and the hadronic shower in the final state are completely determined by the neutrino energy $E_ {\nu}$ and the inelasticity $Y$. Inversely, the measurement of the energies of the produced lepton and hadronic shower can be used to reconstruct the energy of the incoming neutrino, which is one of the main goals of the neutrino telescopes.
 
The average inelasticity  in a neutrino interaction at the detector is defined by 
\begin{equation}
\langle Y(E_{\nu}) \rangle = \frac{\int dY Y \,  \frac{d\sigma_{\nu N}}{dY} }{\int dY  \frac{d\sigma_{\nu N}}{dY}  } = 
\frac{\int dY  Y \, \int dx \, \frac{\partial^2 \sigma_{\nu N}}{\partial x \partial Y}
  }{\int  dY \, \int dx \, \frac{\partial^2 \sigma_{\nu N}}{\partial x \partial Y}
    }\,\,,
\label{Eq:Ymed}
\end{equation}
where the the double differential cross-section for a CC neutrino - hadron interaction is given by (See e.g. Refs. \cite{Kretzer:2002fr,book})
\begin{eqnarray}
\frac{\partial^{2}\sigma_{\nu N}}{\partial x\partial Y} &=& \frac{G^{2}_{F}m_{N}E_{\nu}}{\pi}\left(\frac{M^{2}_{W}}{Q^{2}+M^{2}_{W}} \right)^{2}\left\{    
\left(xY^{2} + \frac{m^{2}_{l}Y}{2E_{\nu}m_{N}}   \right)F_{1}(x,Q^{2})\right.\nonumber \\
&+&\left.  \left[ \left( 1 - \frac{m^{2}_{l}}{4E^{2}_{\nu}} \right)  - \left( 1+\frac{m_{N}x}{2E_{\nu}}\right)Y  \right] F_{2}(x,Q^{2}) 
\right.  \nonumber \\
&+&\left.  \left[ xY\left( 1 - \frac{Y}{2} \right)  - \frac{  m^{2}_{l}Y  }{4E^{2}_{\nu}m_{N}}  \right] F_{3} (x,Q^{2})
 \right\}\,\,,
 \label{Eq:double}
\end{eqnarray}
where $m_l$ is the mass of the lepton produced in the final state, which we  keep  in order to taken into account the effects due to the tau mass,  and we have assumed that the Albright-Jarlskog relations are valid, which is a reasonable approximation for the energies of interest in this analysis \cite{Reno:2021hrj}. For a CC antineutrino - hadron interaction one has that the signal of the last line in Eq. (\ref{Eq:double}) is negative.
Moreover, the functions $F_i$ are the nucleon structure functions that are determined by the underlying structure of the target. As a consequence,  the energy behaviour of 
$\langle Y(E_{\nu}) \rangle$ is strongly dependent on the QCD dynamics at high energies (See, e.g. Refs. \cite{Parente,Goncalves:2013kva,Klein:2020nuk}), which is expected to be modified by nonlinear effects that are predicted  to contribute
at high energies due to the high partonic density present at small values of the Bjorken - $x$ variable. Usually, the neutrino predictions are derived using the solutions of the  linear
DGLAP evolution equations \cite{dglap}. Such equation only considers  the mechanism $g \rightarrow gg$, which populates the transverse space with a large number of small size gluons per unit of rapidity (the transverse size of a gluon with momentum $k_T$ is proportional to $1/k_T$). Such approximation becomes unrealistic for small $k_T$ and large energies, where the produced gluons overlap and the fusion process, $gg \rightarrow g$, becomes equally important. In this regime, DGLAP evolution must be generalized to take into account nonlinear effects, which reduce  the increasing  of the gluon distribution and restore the unitarity of the cross-section \cite{hdqcd}. In recent years, several authors have discussed the impact of these nonlinear effects on $\sigma_{\nu  N}$ and related quantities using the color dipole picture and the Color Glass Condensate (CGC) formalism (See, e.g. Refs. \cite{Goncalves:2010ay,Goncalves:2015fua,Albacete:2015zra,Arguelles:2015wba,bgr18,Goncalves:2021gcu}). An alternative is
to consider the BBMT approach proposed in 
Refs. \cite{Berger:2007ic,Block:2010ud,Block:2013mia,Block:2013nia},  which takes into account the unitarity (saturation) effects at all orders. The main assumption in the BBMT approach for neutrino - hadron interactions is that the growth of the proton structure function is limited by the Froissart bound at high hadronic energies, giving an $\ln^2(1/x)$ bound on $F_2$ as Bjorken $x\rightarrow 0$, which implies 
an exact bound of $\ln^3 E_{\nu}$ for the  $\nu N$ scattering \cite{Illarionov:2011wc}. It is important to emphasize that such approach is able to describe the combined HERA data  \cite{Block:2013mia}. As demonstrated in Ref. \cite{Goncalves:2021gcu}, the BBMT model    
implies a strong reduction of the cross-section at large neutrino energies in comparison to the DGLAP and CGC predictions, and can be considered a lower bound for $\sigma_{\nu  N}$.  In what follows, we will compare the BBMT predictions with those derived using the solutions of the DGLAP evolution equations obtained in Ref. \cite{ct14}.

\begin{figure}[t]
\begin{center}
\begin{tabular}{cc}
    \includegraphics[scale=0.32]{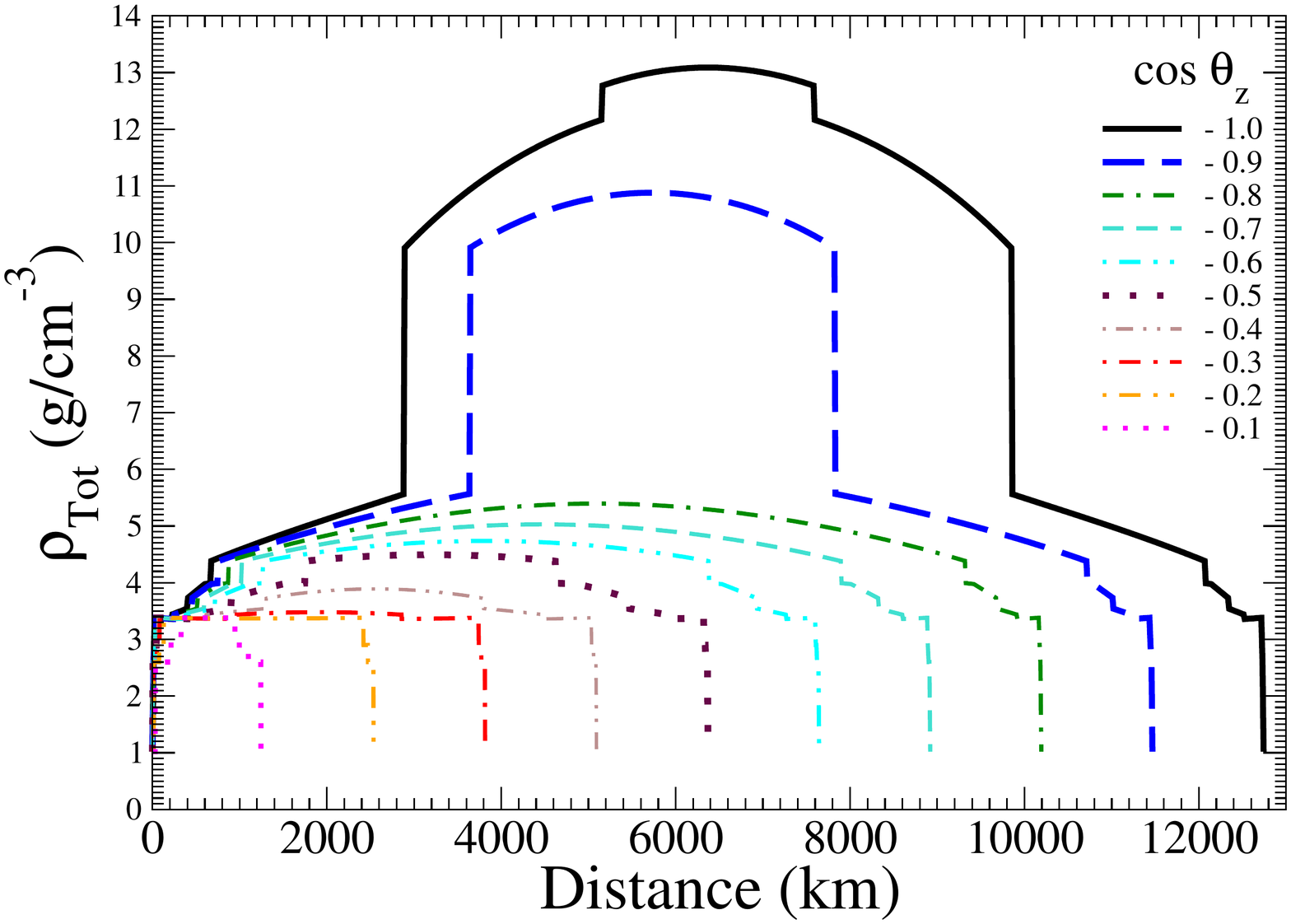} & \includegraphics[scale=0.32]{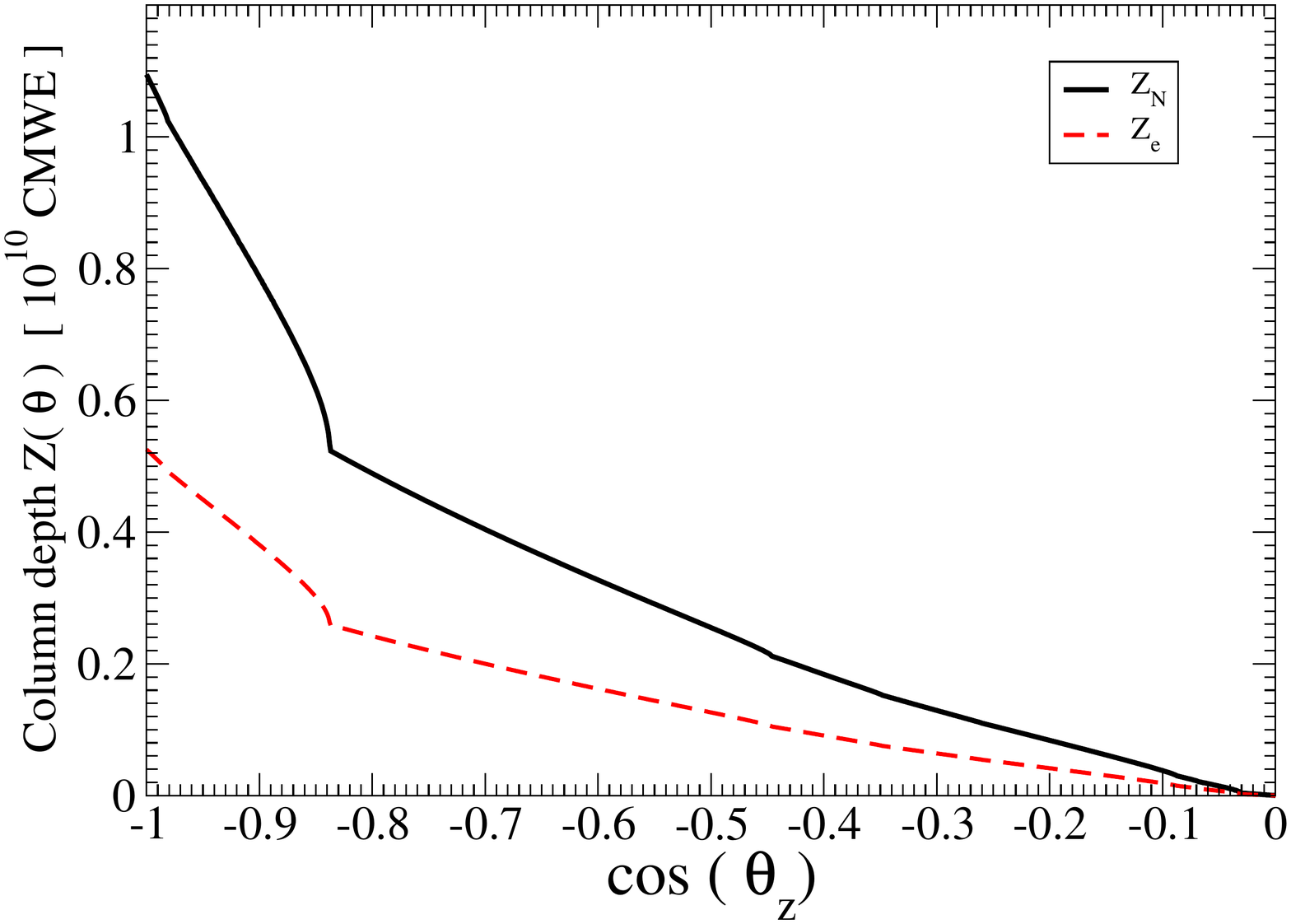} \\
    (a) & (b)
\end{tabular}
   \caption{(a) Relation between the total distance traveled and the total density felt by the neutrinos while crossing the Earth, for the different incoming neutrino directions which are indicated by the values of $\cos{\theta_{z}}$ quoted in the plot. (b) The thickness of the Earth, which is defined in Eq.  (\ref{Eq:zfun}), as function of  $\cos{\theta_{z}}$ for the nucleons ($Z_{N}$) and electron  ($Z_{e}$) targets.  CMWE  stands for  centimeters of water equivalent.  Our results are based on the PREM model \cite{PREM}.   }
    \label{Fig:Rho}
\end{center}
\end{figure}

Moreover, the treatment of the neutrino - nucleon interaction has a direct impact on the 
 probability of neutrino interaction while crossing the Earth, which is defined by  $
P^N_{Shad}(E_{\nu},\theta_{z})=\exp\left\{- \frac{ \,Z_{N}(\theta_{z})}{{\cal{L}}^{N}_{int}}\right\} $,
where $\theta_{z}$ is the zenith angle [See Fig. \ref{Fig:DIS} (b)] and 
${\cal{L}}^{N}_{int} ={1}/({N_{A}\sigma_{\nu N}(E_{\nu})}) $ is the interaction length with nucleons ($N_A$ is the Avogadro's number) \cite{gqrs96}. Besides,  $Z_{N}(\theta_{z})$ is the total amount of matter that neutrinos  feel while it crosses the Earth, which is defined by
\begin{equation}
Z_{N}(\theta_{z})=\int^{r(\theta_z)}_{0}\rho_{N}(r)\,dr,
\label{Eq:zfun}
\end{equation}
where $r(\theta_{z}) = -2 \,R_{\oplus} \cos\theta_{z}$ is the total distance travelled by neutrinos,  and $\rho_{N}(r)[g~cm^{-3}]$  is the density profile of the Earth. In our analysis, we will assume the PREM model for this quantity \cite{PREM}. In order  to illustrate the relation between the incoming neutrino direction and the medium density that they effectively cross, in Fig. \ref{Fig:Rho} (a) we show how the total path and density of the medium that neutrinos crosses are highly sensitive to the incoming direction. For instance, notice that only in the interval of  $\cos{ \theta_{z}} \lesssim -0.84 $ neutrinos enter  the Earth's core, and the inner core is only accessible for $\cos{ \theta_{z}} \rightarrow -1.0 $. In Fig. \ref{Fig:Rho} (b) we show  how this dependence reflects in the total amount of matter that neutrinos go through.  Indeed, the effect of crossing all the Earth's core implies  a factor $\approx 2$ in $Z_N(\theta_z)$ in comparison with the case of $\cos{ \theta_{z}} = -0.8 $, where the path is large ($\approx 10^{4}$ km), but neutrinos do not enter  the core.  For comparison, we also present in Fig. \ref{Fig:Rho} (b) the predictions for $Z_e$, which is associated with neutrino - electron interactions during the propagation through the Earth, which has been estimated following Ref. \cite{Goncalves:2021gcu}.

\begin{figure}[t]
    \centering
    \includegraphics[scale=0.45]{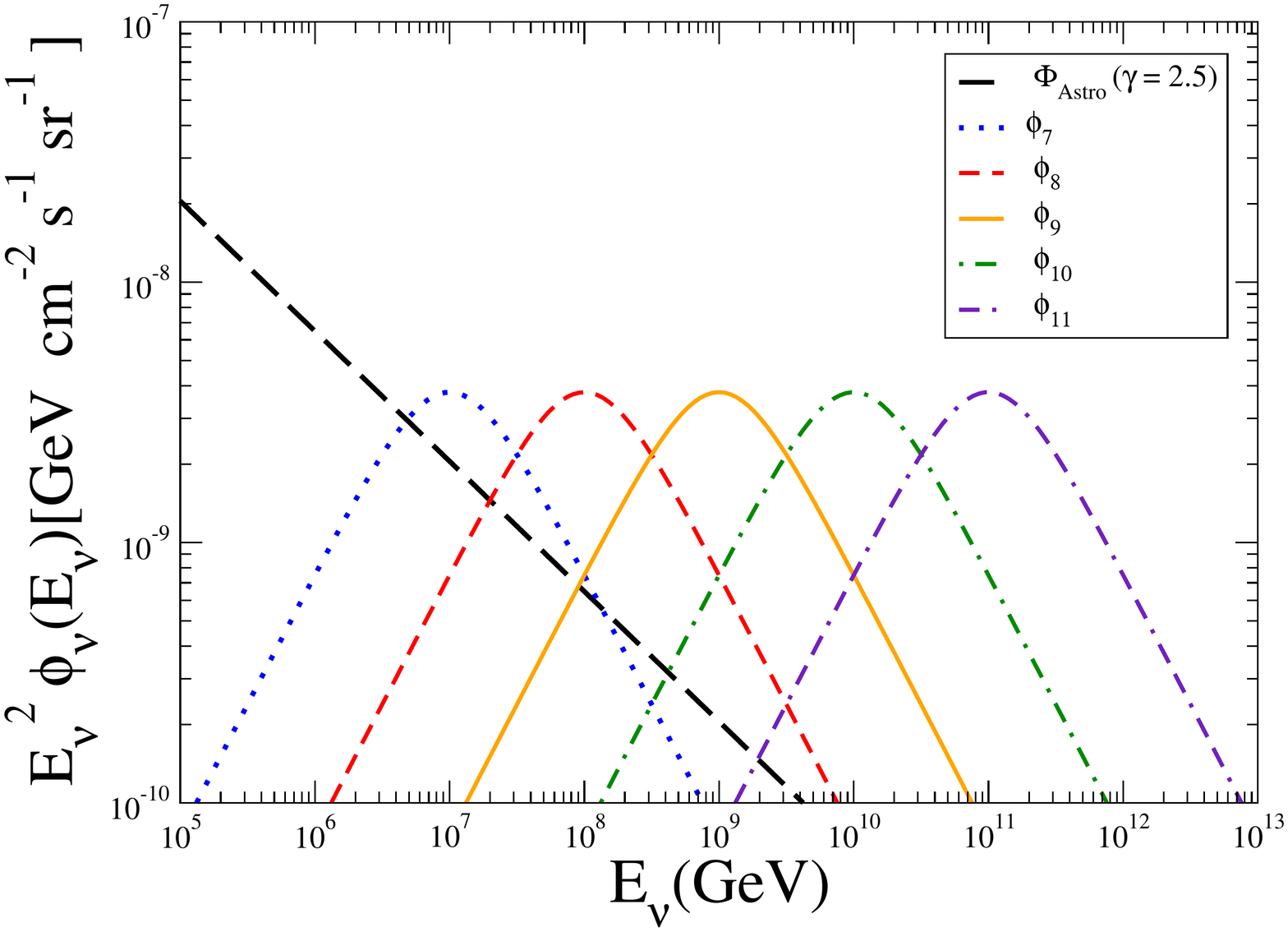}\
    \caption{Comparison between the  standard astrophysical flux  $\Phi_{astro}(E_{\nu})$,  given by Eq. (\ref{Eq:flux}) with $\gamma = 2.5$ and $\Phi_{0} =2.0 f.u.$,  and the {Super-Glashow} astrophysical  neutrino fluxes, $\phi_{j}(E_{\nu})$, which are given by Eq. (\ref{Eq:fluxSuperGlashow}).  For the { Super-Glashow} astrophysical  neutrino fluxes, one has imposed an equal peak normalization. }
    \label{Fig:SuperGlashow}
\end{figure}

Indeed, $\sigma_{\nu  N}$ and $P^N_{Shad}(E_{\nu},\theta_{z})$ are two of the main ingredients to estimate  the differential rate of astrophysical neutrino - induced events in neutrino telescopes, which is given by 
\begin{eqnarray}
\frac{dN_{events}}{d(E_{vis}) d\Omega}       &=& T \sum_{\alpha}N_{eff,\alpha}(E_{\nu})\times \Phi_{\nu_{\alpha}}(E_{\nu})\times \sigma_{\nu_{\alpha}N}(E_{\nu}) \times P^N_{Shad}(E_{\nu},\theta_{z}) ,
\label{Eq:neve}
\end{eqnarray}
where $T$ is the time of data taken, $N_{eff,\alpha}(E_{\nu})$ is  the effective number of scattering targets, and $\Phi_{\nu_{\alpha}}$ is the astrophysical neutrino flux for
a neutrino of  flavor $\alpha$. The third main ingredient is the astrophysical neutrino flux $\Phi_{\nu_{\alpha}}(E_{\nu})$, which origin   is still a theme of intense debate, but  so far consistent  with results expected from extra-galactic sources,
presenting isotropy and no correlation with the galactic plane. In our analysis, we will assume  the same astrophysical neutrino flux for the three 
neutrino flavors and that the astrophysical flux is given by \cite{IceCube:2017zho,IceCube:2020wum,IceCube:2021uhz,IceCube:2020acn}
\begin{eqnarray}
\Phi_{astro}(E_{\nu})=   \sum_{\alpha} \Phi_{\nu_{\alpha}}(E_{\nu}) =  {\Phi_{0}} \times \left(\frac{E_{\nu}}{100~TeV}\right)^{-\gamma} (f.u.).
\label{Eq:flux}
\end{eqnarray}
  In the next Section, we will estimate the distribution of neutrino events at the IceCube assuming different assumptions for the QCD dynamics and for the inelasticity, and we will determine the best estimates for $\Phi_{0}$ and $\gamma$  using a maximum likelihood fit by the comparison
of our predictions with the distribution of observed events. 

Finally, following Ref. \cite{Kistler:2016ask},  we will also consider the possibility that a new (still hypothetical)  contribution for the astrophysical neutrino flux peaks at  energies larger than that  characteristic of the {Glashow Resonance}, $E^{res}_{\nu}\approx 6.3$ PeV. In this case, we will assume that 
total astrophysical neutrino flux is expressed by 
\begin{equation}
\Phi^{tot}_{astro}(E_{\nu}) = \Phi_{astro}(E_{\nu})\mbox{[Eq. (\ref{Eq:flux})]} + \phi_{j}(E_{\nu}) \,\,\,,   
\label{Eq:PhiTot}
\end{equation}
where the {\it Super-Glashow} flux $\phi_{j}(E_{\nu})$ is parameterized as follows \cite{Kistler:2016ask}
\begin{equation}
\phi_{j}(E_{\nu}) = \phi_{0j} \times \left[ \left( \frac{E_{\nu}}{E_{0j}} \right)^{\alpha\eta} +
\left( \frac{E_{\nu}}{E_{0j}} \right)^{\beta\eta}\right]^{1/\eta}\,\,. 
\label{Eq:fluxSuperGlashow}
\end{equation}
Here $j$ denotes the energy in which the flux is maximum,  $E_{0j} = 10^{j}$ GeV, and  $\phi_{0j}$ is the  corresponding flux normalization. In our analysis, we will assume $\eta = -1$, $\alpha = -1$, and $\beta = -3$, in order to include some desired characteristics of the source and cosmic evolution, and consider the following possibilities: $j= 7, 8, 9, 10$ and 11.   As discussed in the Introduction, the current data are quite well described disregarding the presence of a new component. However, the possibility of a new component that peaks for energies not yet covered by the current experiments cannot be discarded, which makes this topic a theme of intense debate. 
  In Fig. \ref{Fig:SuperGlashow} we illustrate the energy range where the hypothetical {Super-Glashow} fluxes are expected to become  important, as well as its energy dependence.  Notice that, if present, the {Super-Glashow} fluxes will enhance the expected number of neutrino events precisely in the energy range  where the nonlinear QCD effects become non-negligible.

% by assuming {\bf \color{red}  In Ref. \cite{Kistler:2016ask}, the authors shown that these fluxes are more likely to produce the reported PeV track-like signals at the IceCube observatory than the astrophysical neutrino flux given by Eq. (\ref{Eq:flux}). The reason is simply that to produce such high-energy neutrino events, the  flux following   Eq. (\ref{Eq:flux}) will require higher values of the flux normalization parameter, $\Phi_{0}$, or smaller values of the power index parameter, $\gamma$. Both features  imply a higher number of events at lower energies, which are not present in the data.:} {\bf Sugestão: Substituir por:} {\color{blue} In Ref. \cite{Kistler:2016ask}, the authors shown that these fluxes are more likely  to generate the ultrahigh-energy neutrino events at the observatory than the standard astrophysical flux, which is centered at $E_{0}=10^{5}$ GeV. The reason is simply that these ultrahigh-energy fluxes are centered at higher neutrino energies than $\Phi_{astro}$} .  

\begin{figure}[t]
\begin{center}
\begin{tabular}{cc}
\includegraphics[scale=0.36]{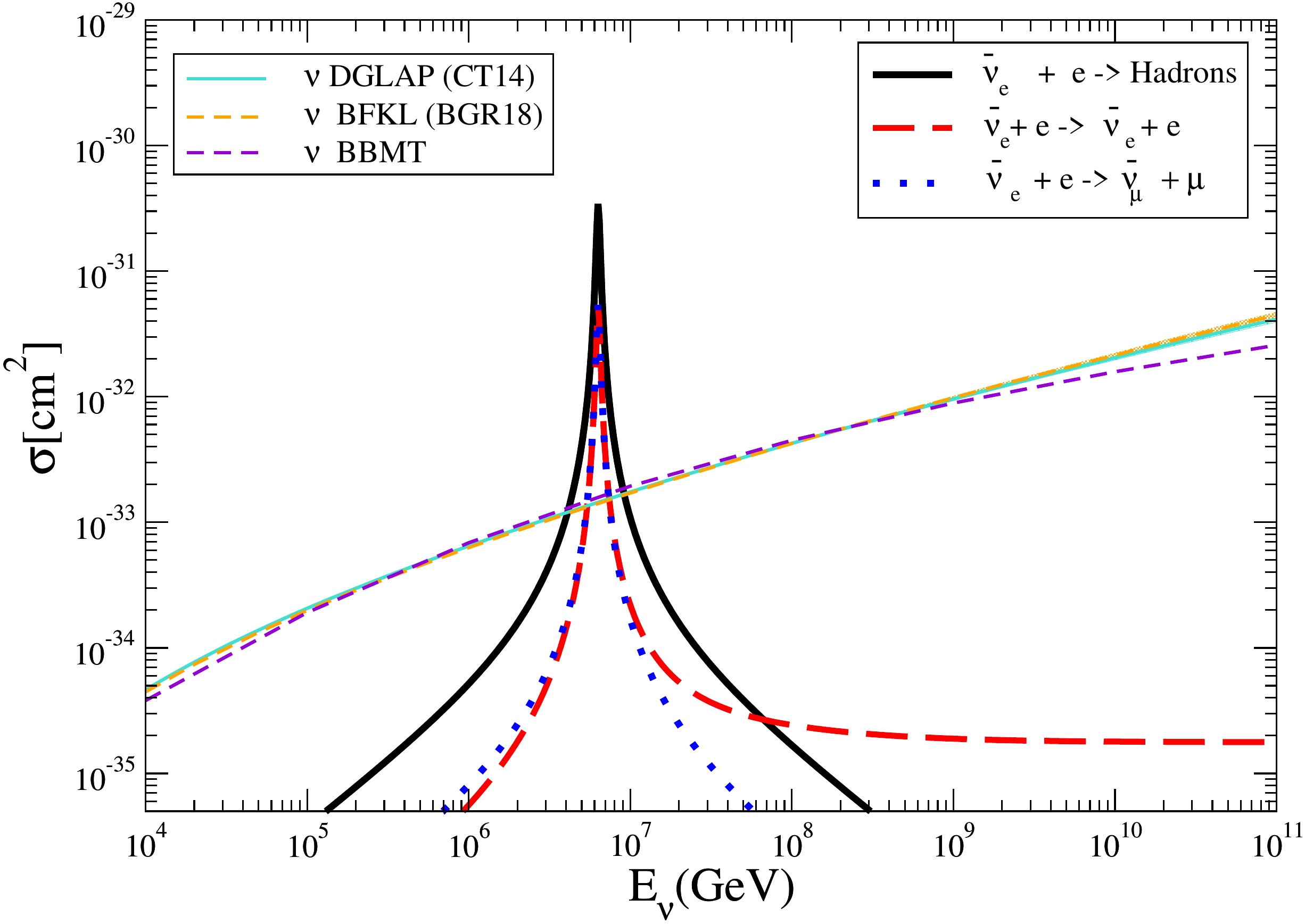} &  \includegraphics[scale=0.36]{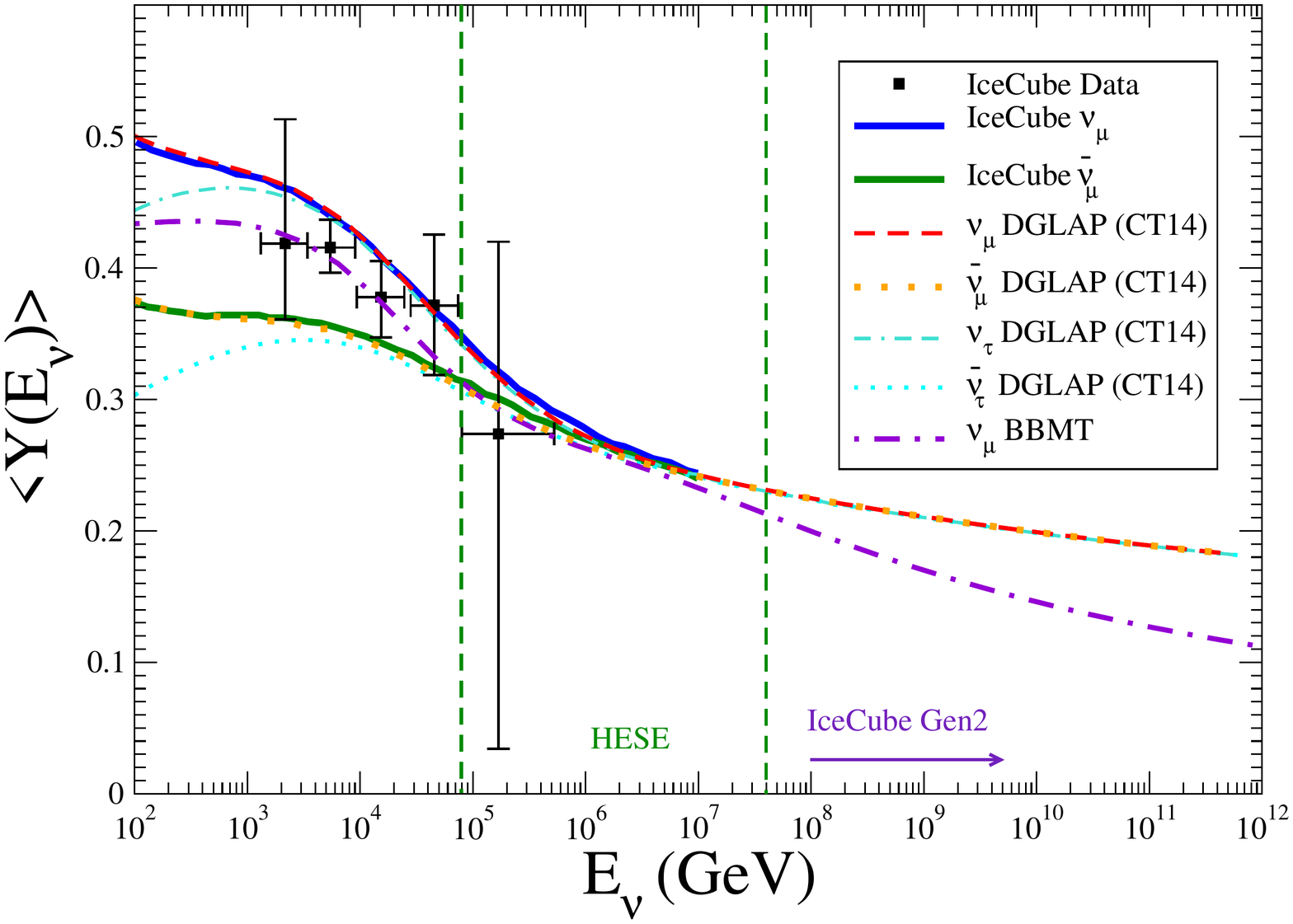} \\
(a) & (b)\\
\end{tabular}
\caption{(a) Energy dependence of the neutrino - target cross-section considering different models for the QCD dynamics; (b) Predictions  for the energy dependence of the average inelasticity. For comparison, the   data  and theoretical predictions from \cite{IceCube:2018pgc} (solid lines) are also presented.}
\label{Fig:CS}
\end{center}
\end{figure}

\section{Results}
\label{Sec:res}

Initially, let us present in Fig. \ref{Fig:CS} (a), for completeness of our analysis, the predictions for the energy dependence of the neutrino - target cross-section. For the CC neutrino - nucleon case,
we present the DGLAP prediction,  estimated using  
 the CT14 parameterization \cite{ct14} and denoted by DGLAP (CT14) hereafter, as well as the BBMT one. For comparison, we also present the prediction derived in Ref. \cite{bgr18}, denoted BFKL (BGR18) in the figure, which has been obtained using the framework of collinear factorization at NNLO and  taking into account  the small - $x$ BFKL resummation up to next - to - leading logarithmic (NNLx) accuracy.  
 As it can be seen, the DGLAP (CT14) and BFKL (BGR18)  results are similar in the IceCube energy range, but its central predictions are slightly distinct for larger energies. In contrast, 
the BBMT result is similar to the DGLAP and BFKL predictions  in the IceCube energy range, but implies a strong reduction of the cross-section at larger neutrino energies. 
In particular, the predictions for large neutrino energies can differ by a factor $\ge 2$ depending on the approach assumed to treat the
QCD dynamics. Our results for the antineutrino - electron 
cross-section are also presented in Fig. \ref{Fig:CS} (a) taking into account the presence of the Glashow resonance. These results indicate that the  $\bar{\nu}_e e$ scattering  becomes equal or greater
than CC neutrino-nucleon cross-section in the  energy range characterized by  $10^6$ GeV $\le E_{\nu}\le 2 \times 10^7$ GeV. 

In Fig. \ref{Fig:CS} (b) we present the corresponding predictions for the energy dependence of the average inelasticity $\langle Y \rangle$. For comparison, we also present the data and   predictions from  Ref. \cite{IceCube:2018pgc}.
Assuming an incoming muon neutrino and antineutrino flux and the DGLAP approach, one has that our results 
agree with those presented in Ref. \cite{IceCube:2018pgc} in  the IceCube energy domain. 
In addition, in the energy range of the {High Energy Sample of Events (HESE)}, one has that $
 0.23 \le \langle Y(E_{\nu}) \rangle \le 0.35$, which implies that the assumption  $\langle Y \rangle = 0.35$, sometimes present in the literature, overestimates the average inelasticity by a factor of the order of $52\%$ in the limit of ultrahigh neutrino energy.
In order to estimate the impact of the tau mass, we also present our predictions for an incoming $\nu_{\tau}$ and $\bar{\nu}_{\tau}$ flux. Our results indicate  that it cannot be disregarded for  $E_{\nu} \leqslant 10^{4}$ GeV.
Finally, let's analyze the impact of the QCD dynamics on $\langle Y \rangle$ by comparing the DGLAP (CT14) and BBMT predictions. Indeed, the BBMT model  implies  a systematically lower  $\langle Y(E_{\nu}) \rangle$ for all the energy range considered. However, the difference between the predictions is not large in the energy range of the current data, being  of the order of $5\%$ for $E_{\nu} = 10^{7}$ GeV. On the other hand, our results indicate  that they could have an important role in the near future, when the increase in the number of events will  improve the detector sensitivities to neutrino events above the Glashow resonance.  For instance, the BBMT model implies a reduction of $55\%$  for $E_{\nu} = 10^{11}$ GeV. In comparison to the approximation $\langle Y \rangle = 0.35$, it implies an average inelasticity that is smaller by a factor of 3, which indicates that the energy evolution of the average inelasticity cannot be disregarded in the forthcoming neutrino telescopes.

\begin{figure}[t]
\begin{center}
%\begin{tabular}{cc}
\includegraphics[scale=0.45]{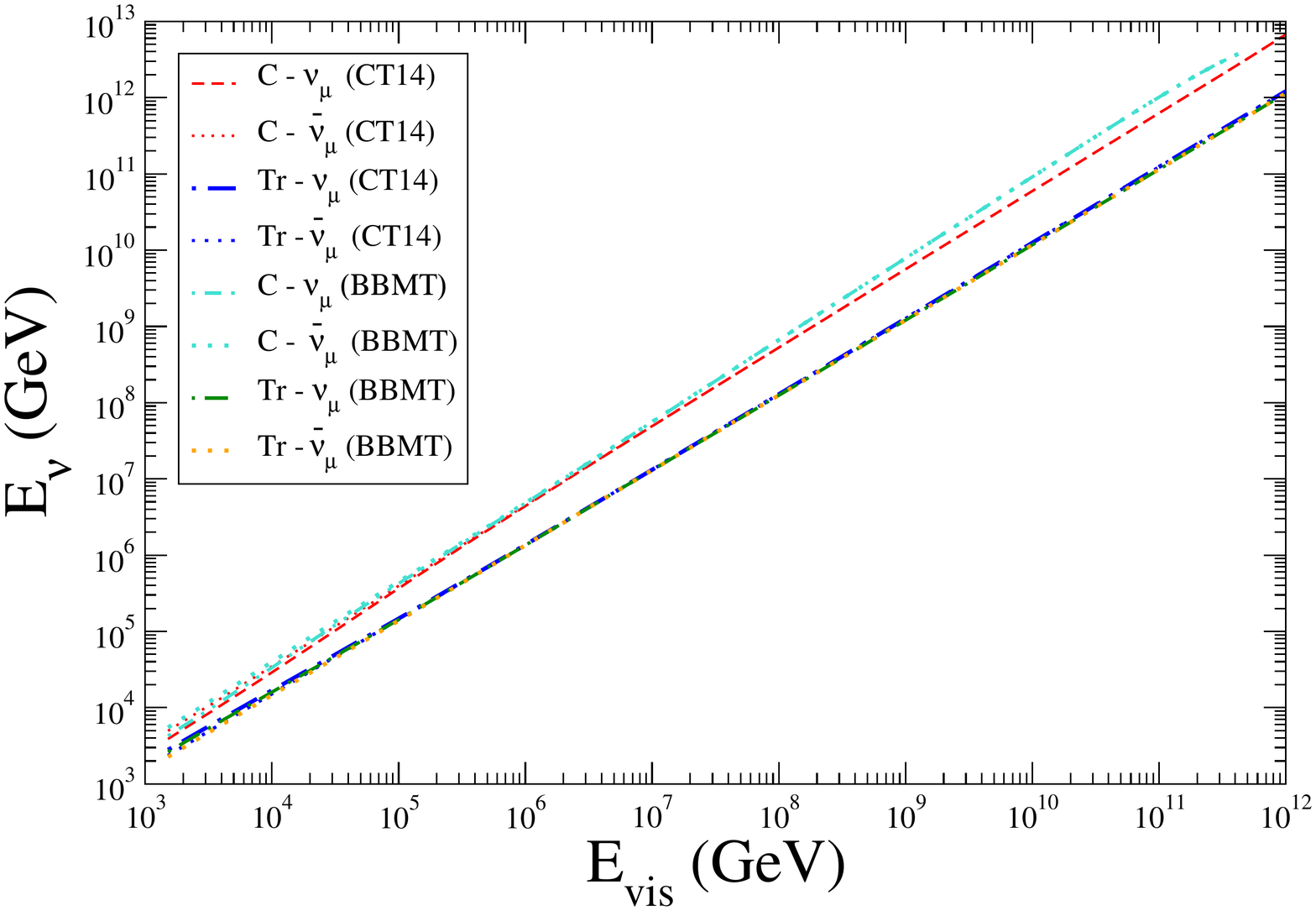} 
\caption{ The impact of the average inelasticity in the relation 
between the incoming neutrino energy and the visible energy in the detector for the (anti)muon neutrino case. The label {\it C} ({\it Tr}) refers to {\it cascades} ({\it muon tracks}).}
\label{Fig:Evis}
\end{center}
\end{figure}

Concerning the consequences for the energy of the products of the neutrino interaction, our results indicate that  the lower and higher energy limits of the HESE sample are, respectively, $E_{\nu}=1.5 E_{l}$ and $E_{\nu} = 1.2 E_{l}$. It follows that $E_{l}$ and $E_{\nu}$ are approximately of the same order of magnitude in the energy range considered here, especially in the UHE limit. In addition,  the ratio between the neutrino energy and the energy transferred to the target at the  inferior (superior)  limit of the HESE sample, will be    $E_{\nu} = 2.9 E_{X}$ ($E_{\nu} = 5.0 E_{X}$). In Fig.  \ref{Fig:Evis}, we analyze  the impact of the nonlinear effects on the relation between the  neutrino energy and  the energy of the products of its interaction, i.e.  muon tracks (Tr) and hadronic cascades (C) produced at the  vertex of 
the (anti)muon neutrino interaction. For $~E_{vis}=10^{11}$ GeV  we have $\langle Y(CT14)\rangle = 0.17 $ and  $\langle Y(BBMT) \rangle = 0.13$, which implies the ratios
\begin{equation}
 R_{Y,C}=\frac{\langle Y(BBMT) \rangle}{\langle Y(CT14)\rangle}=0.76 ~~;~~~~~~~R_{Y,Tr}=\frac{1-\langle Y(BBMT) \rangle}{1-\langle Y(CT14)\rangle}=1.05.
\label{Eq:RY}
 \end{equation}
Therefore,  we have an effect of $\approx 25\%$ in the transferred energy from the neutrino to the hadronic shower, but of only $\approx5\%$ at the muon track. Notice that if the resulting charged lepton produces a track in the detector, as a muon or a long-lived tau, the  effects due to energy loss are non-negligible and modify the relation between the incoming neutrino energy and the effective amount of energy deposited inside the detector, in such way that higher values of neutrino energy should be necessary to result in the same visible energy.  Such topic deserves  more detailed analysis, as that performed in  \cite{Kistler:2013my}. However, we emphasize that the results presented here are still valid since they refer to the primary vertex of the interaction.

\begin{figure}[th]
\begin{center}
\begin{tabular}{cc}
\includegraphics[scale=0.35]{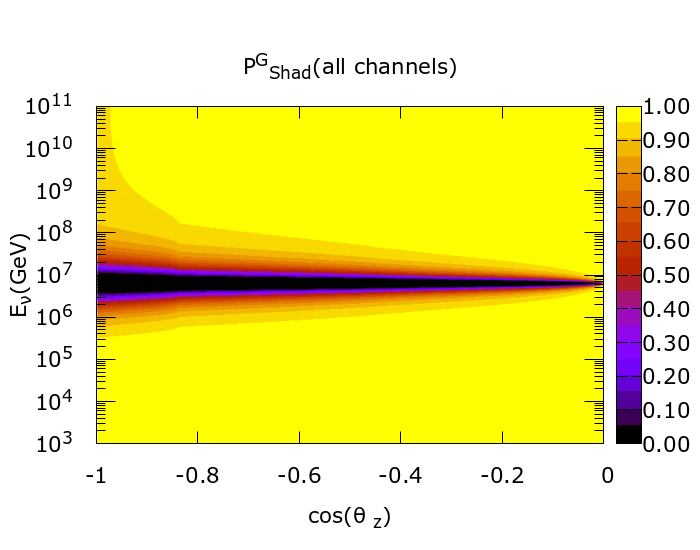} &
%\hspace{-0.25cm}\includegraphics[scale=0.325]{Pshad_CT14_2022.png} \\
\includegraphics[scale=0.35]{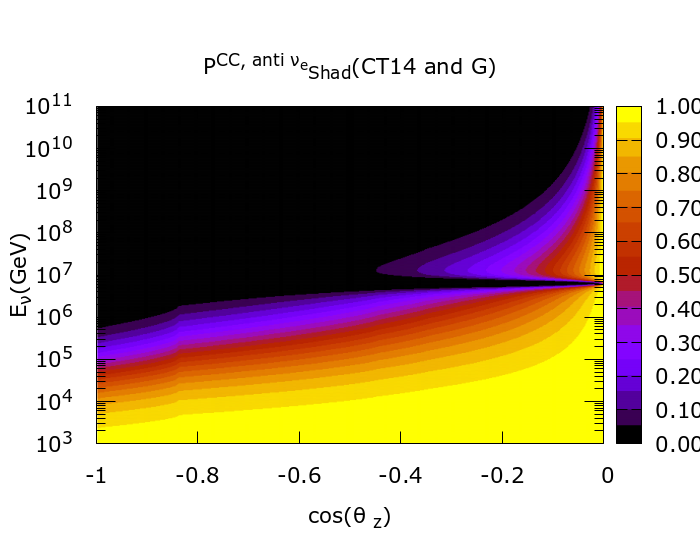} \\
(a) & (b)
\end{tabular} 
%\\
%\includegraphics[scale=0.35]{Pshad_nuanu_CT14_2022.png}
\caption{The energy and angular dependencies of the probability of antineutrino absorption by the Earth, $P_{shad}(E_{\nu},\theta_{z})$ for (a) the Glashow resonance and (b) the sum of the contributions associated with the Glashow resonance and antineutrino - hadron cross-section estimated using the DGLAP (CT14) approach.}
\label{Fig:Pshad1}
\end{center}
\end{figure}

\begin{figure}
\begin{center}
\begin{tabular}{cc}
\includegraphics[scale=0.35]{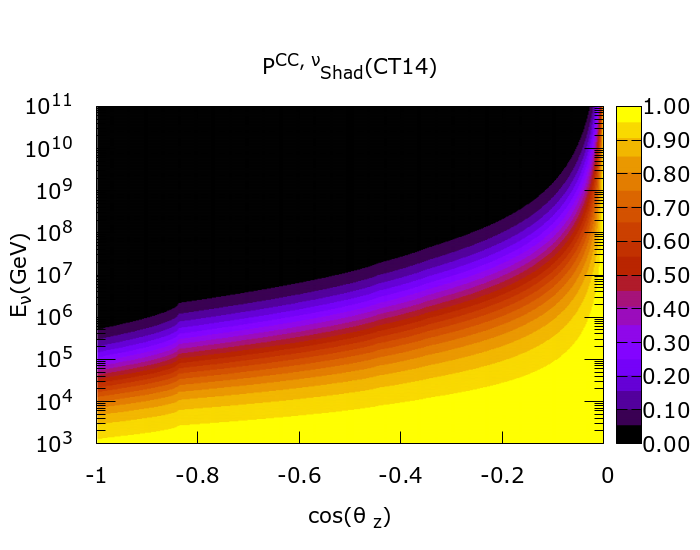} & 
\includegraphics[scale=0.35]{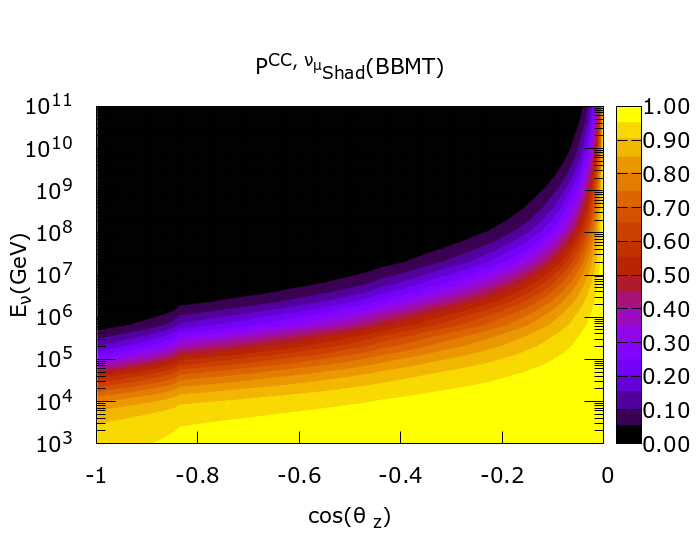} \\
(a) & (b) \\
\includegraphics[scale=0.35]{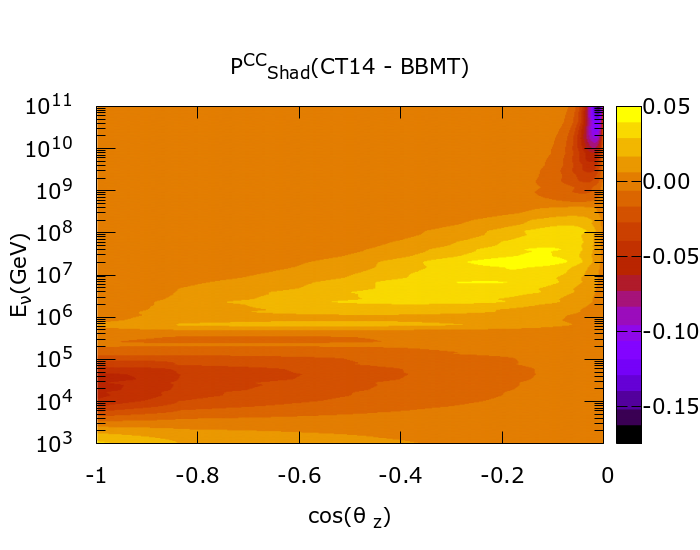} & 
\includegraphics[scale=0.35]{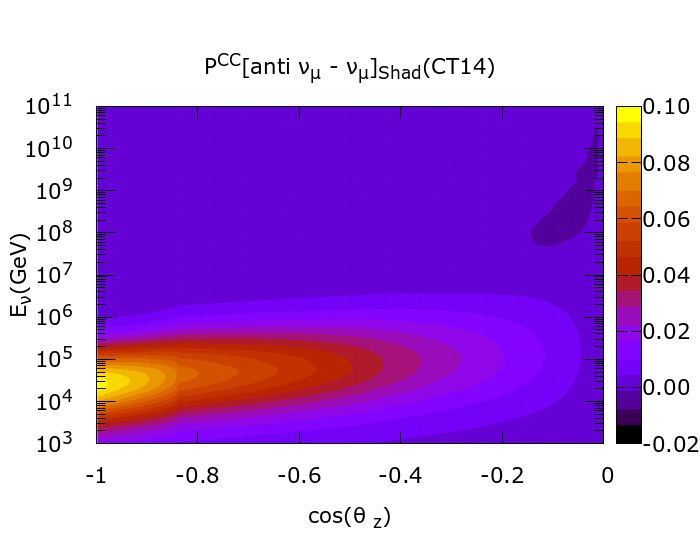} \\
(c) & (d)
\end{tabular}
\caption{ The energy and angular dependencies of the probability of neutrino absorption by the Earth, $P_{shad}(E_{\nu},\theta_{z})$, with $\sigma_{\nu N}$ calculated using the (a) DGLAP (CT14) approach  and the (b) BBMT model; (c) Difference between the DGLAP (CT14) and BBMT predictions; (d) The difference between muon neutrino and antineutrino
probabilities of absorption estimated assuming the DGLAP(CT14) approach.}
\label{Fig:Pshad2}
\end{center}
\end{figure}

Let us now analyze the angular dependence of the  
 probability of neutrino interaction while crossing the Earth. 
Our results   are presented in Figs. \ref{Fig:Pshad1} and \ref{Fig:Pshad2}. The left 
vertical axis show the neutrino energy $E_{\nu}$ and  the horizontal axis is the cosine of the zenith angle, $\cos(\theta_{z})$. The color pattern on the right informs the value
of $P_{Shad}(E_\nu, \theta_z)$. Moreover,  $\cos(\theta_{z})\rightarrow -1$ refers to incoming neutrinos that cross all the  diameter of the Earth ($\approx 12000$km) before reaching
the detector and feel the scattering potential of all the layers of the Earth's interior, including the core [See Fig. \ref{Fig:DIS} (b)]. On the other hand, $\cos(\theta_{z})\rightarrow 0$ 
refers to incoming neutrinos  from the horizon. In this case, neutrinos travel a few hundred kilometers and cross only the Earth's crust.

 Initially, in Fig. \ref{Fig:Pshad1} (a)  we present our results for the  Glashow resonant scattering of electron antineutrinos. As expected from Fig. \ref{Fig:CS} (a), the resonance has its impact limited to a sharp region around the resonant neutrino energy, $E^{res}_{\bar \nu_{e}}\approx 6.3$ PeV.
 When the antineutrino - nucleon CC interactions are taken into account considering the DGLAP(CT14) approach, the absorption probability is strongly modified, as shown in Fig. \ref{Fig:Pshad1} (b),  mainly in the high energy range, where the Earth becomes opaque. Such behaviour is directly associated with the increase in the $\sigma_{\bar{\nu}N}$ with the neutrino energy.  
 As the probability of antineutrino interaction depends on the product between the antineutrino - hadron cross-section
 and the total amount of matter that antineutrinos go through, which is related to the Earth's  density
profile,  $P_{Shad}(E_{\nu},\theta_{z})$ presents a strong correlation between the incoming antineutrino direction and the antineutrino energy. This dependence is clearly seen in Fig. \ref{Fig:Pshad1} (b). For instance, for $\cos(\theta_{z})\rightarrow -1$  the Earth becomes opaque to antineutrinos for   $E_{\nu} \gtrsim 10^{6}$ GeV, while for 
 $\cos(\theta_{z})\rightarrow 0$, even at $E_{\nu} \gtrsim 10^{11}$ GeV, the antineutrino survival probability is non-negligible. The exact form of this dependence is a consequence of 
 how the thickness of the Earth varies with $\cos(\theta_{z})$ (See Fig. \ref{Fig:Rho}).

 In the upper panels of Fig. \ref{Fig:Pshad2} (a) and (b) we present, respectively, the DGLAP(CT14) and BBMT predictions for 
 $P_{Shad}(E_\nu, \theta_z)$ considering a muon neutrino - nucleon CC interaction. One can see that both models predict similar behaviours. In order, to quantify how distinct are these behaviours, in   panel (c) we present the results for the difference between these two results.  One has that the absolute value of the difference  is  $5\% -15\%$ and depends on the incoming neutrino direction. For  $\cos{\theta_{z}}\rightarrow -1$,  the Earth is opaque to the ultrahigh energy neutrinos independently of the model considered for $\sigma_{\nu N}$. Such result is expected due to  the  large value of $Z(\theta_z)$, which implies that the Earth becomes opaque at neutrino energies lower than the necessary to the nonlinear effects associated with the unitarity corrections become  important.  On the other hand, for $\cos{\theta_{z}}\rightarrow 0 $, the total amount of matter crossed by the neutrino is comparatively smaller, such that the neutrino survival probability is significantly larger than zero, even in the UHE  limit. As a consequence, in this regime, the difference in the predictions for  $P_{shad}(E_{\nu},\theta_{z})$  is appreciable and of the order of $15\%$ at $E_{\nu}\approx 10^{11}$ GeV.  For completeness, in  Fig. \ref{Fig:Pshad2} (d), we present   the difference in the probabilities of absorption for the muon antineutrino  and  muon neutrino cases. This is  indicative of the importance of  $F_{3}(x,Q^{2})$, which encodes the differences between quark and antiquark content inside the target nucleon. As we can see, the difference is important at the IceCube energy range, and reaches a maximum of $10\%$ for  $E_{\nu}$ of the order of a few dozens of TeV and $\cos{\theta_{z}}\rightarrow -1$.

\begin{figure}[t]
\begin{center}
\begin{tabular}{cc}
\includegraphics[scale=0.4]{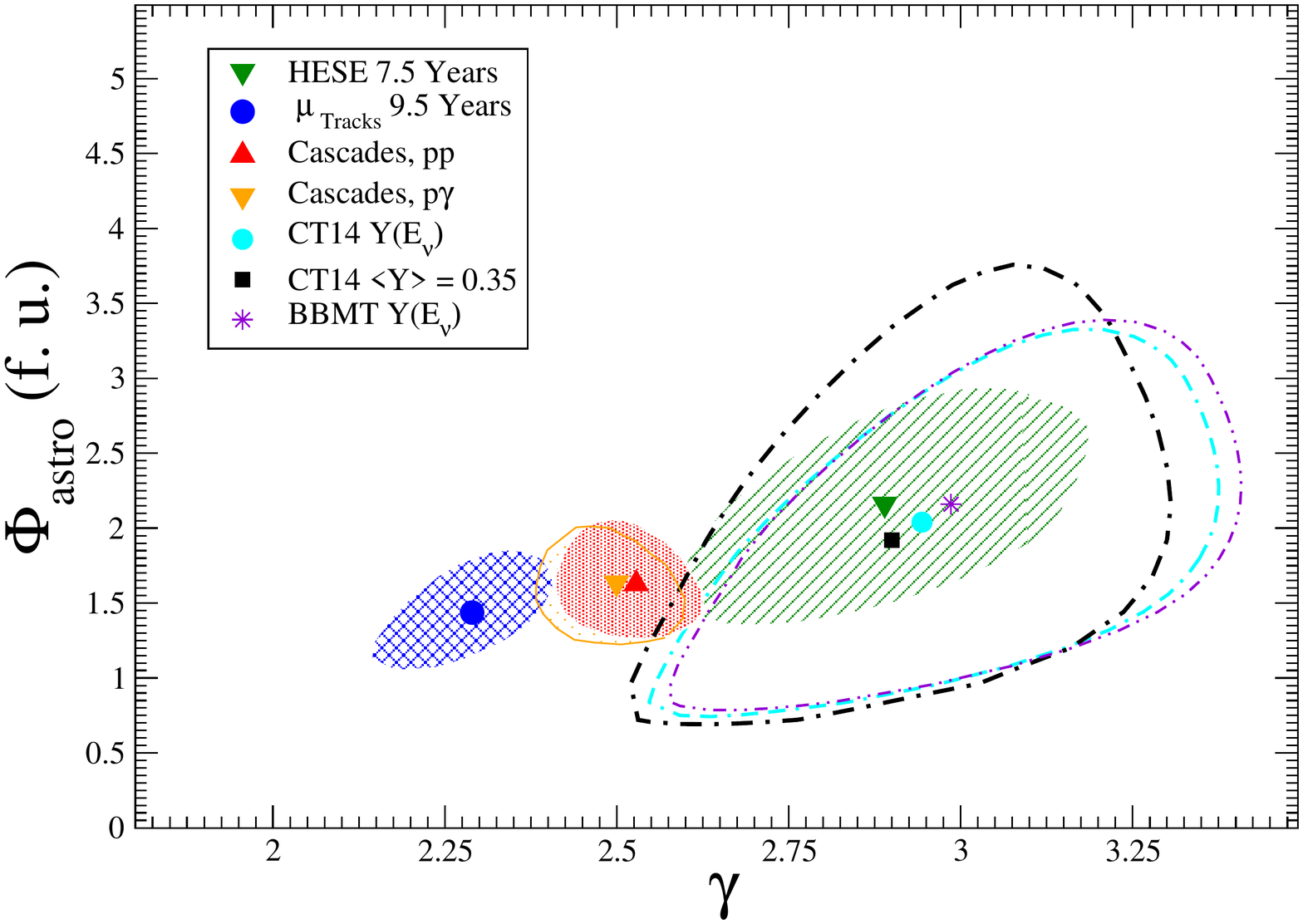}
\end{tabular}
\caption{Results for the {\it likelihood} analysis of the number of neutrino events observed in the IceCube during six years of exposition of the HESE data (dashed lines) considering three different approaches for the treatment of the average inelasticity at the primary neutrino interaction at the
detector. For completeness, the results from Ref. \cite{IceCube:2020acn} are also shown. }
\label{Fig:Like-Y}
\end{center}
\end{figure}

The results derived above for $\langle Y \rangle$ and $P_{shad}(E_{\nu},\theta_{z})$ allow us to estimate the impact of the QCD dynamics in the determination of the astrophysical neutrino flux parameters. For this,  we perform a {\it likelihood} analysis of the IceCube data (For details see Ref. \cite{Goncalves:2021gcu}). In particular, we will 
use the six years of exposure of the { High-Energy Sample of Events (HESE)} and will consider as input the DGLAP (CT14) and BBMT predictions. For comparison, we will also present the results derived assuming 
$\langle Y \rangle = 0.35$.  Our results are presented in Fig. \ref{Fig:Like-Y}.
As expected, both approaches imply an 
equally good description of the data. { However, in contrast with the case of $\langle Y(E_{\nu})  \rangle$, if  $\langle Y \rangle$ is assumed as a constant, the values of $\Phi_0$ and $\gamma$ decrease.  In addition, if the BBMT is assumed in the calculations, the best fit value for the flux 
normalization (spectral index)   is increased by  $ \approx 10\% \,(3\%)$}. Such result indicates that even small modifications in the average inelasticity due to the BBMT interaction model imply  modifications in the description of the number of neutrino events  in the present HESE data.

%   -------------------------------------------------------------------------------------

\begin{figure}[t]
\begin{center}
\begin{tabular}{cc}
\includegraphics[scale=0.31]{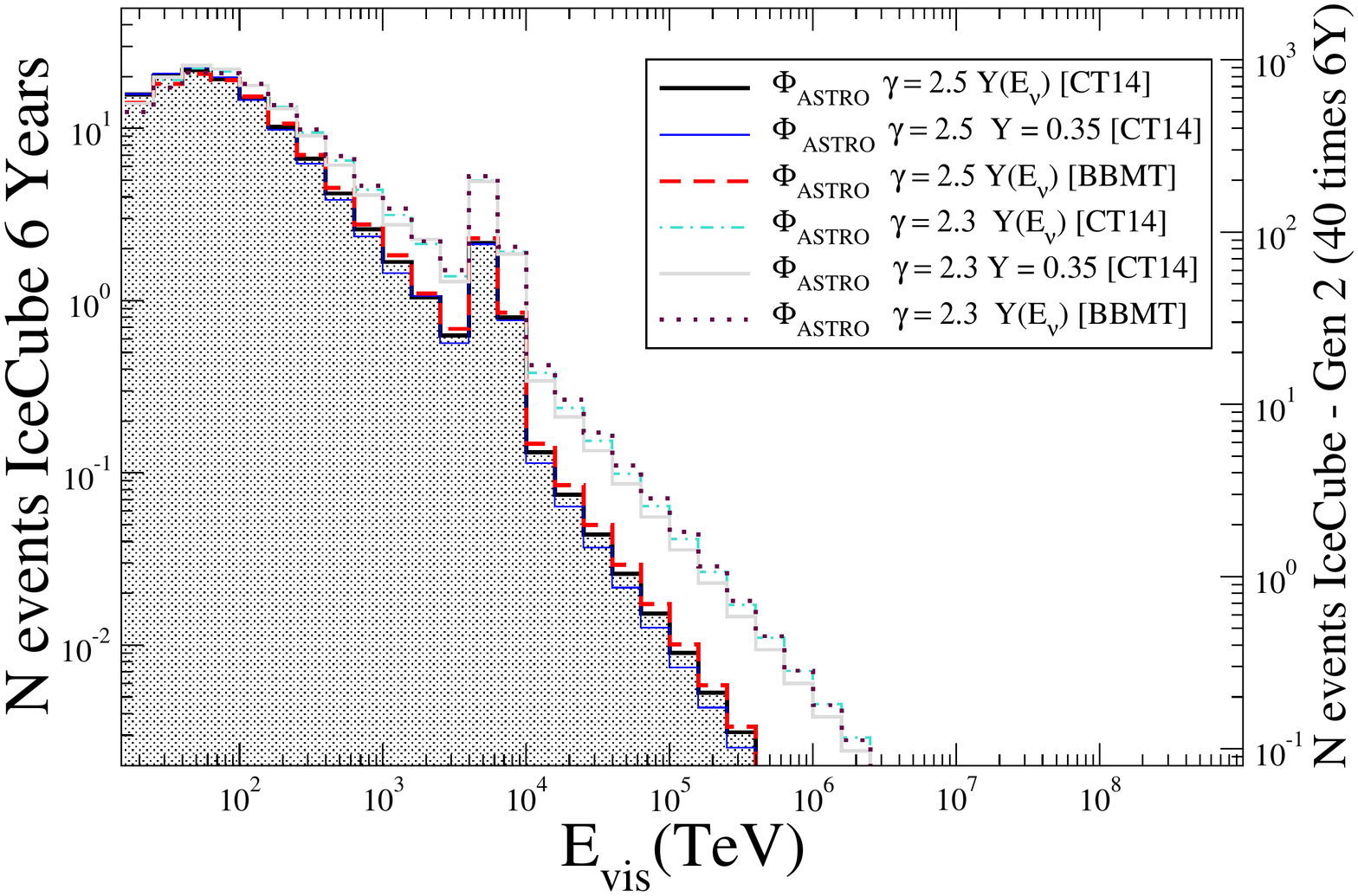} &
\includegraphics[scale=0.31]{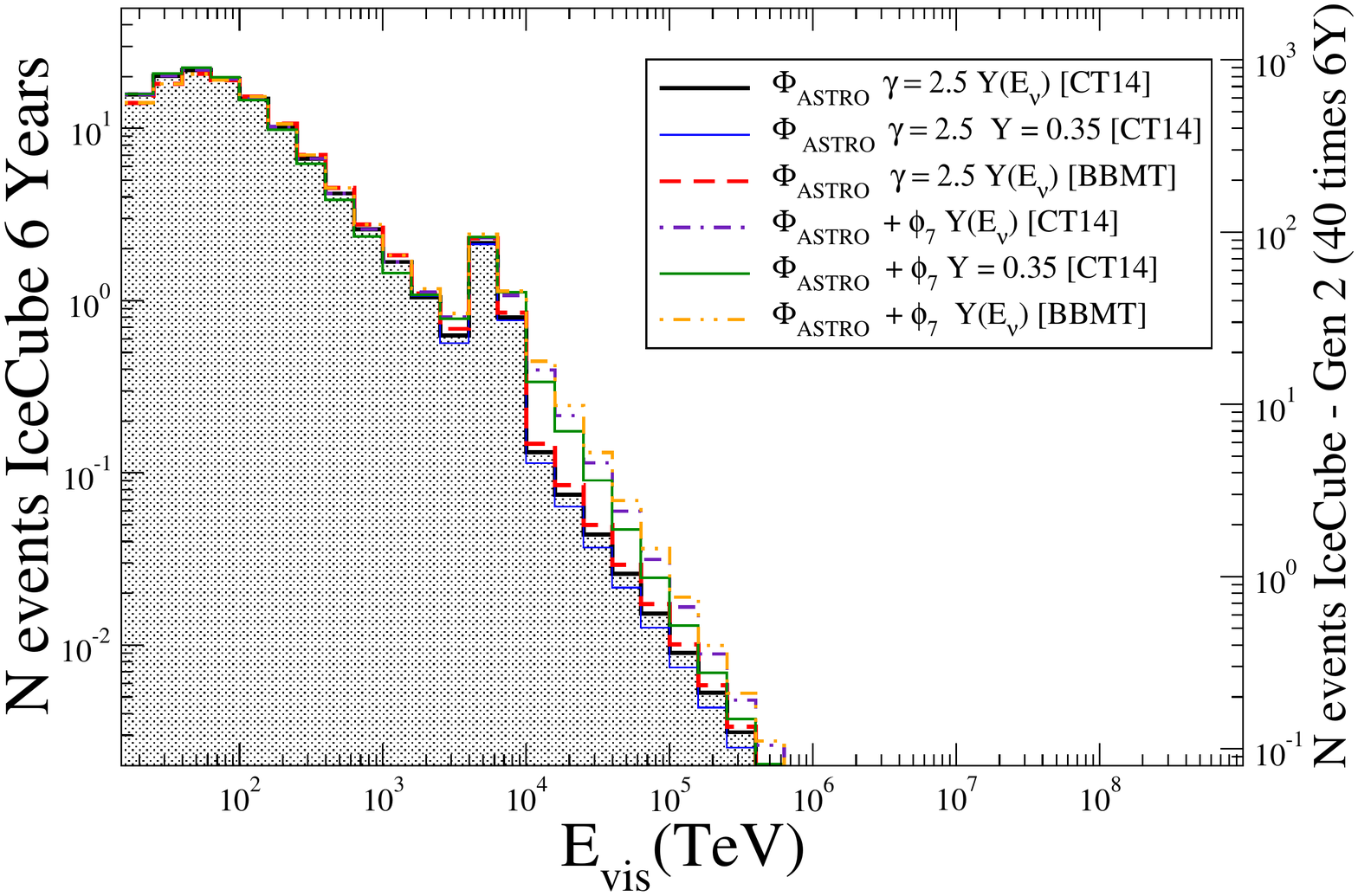} \\
(a) & (b)\\
\includegraphics[scale=0.31]{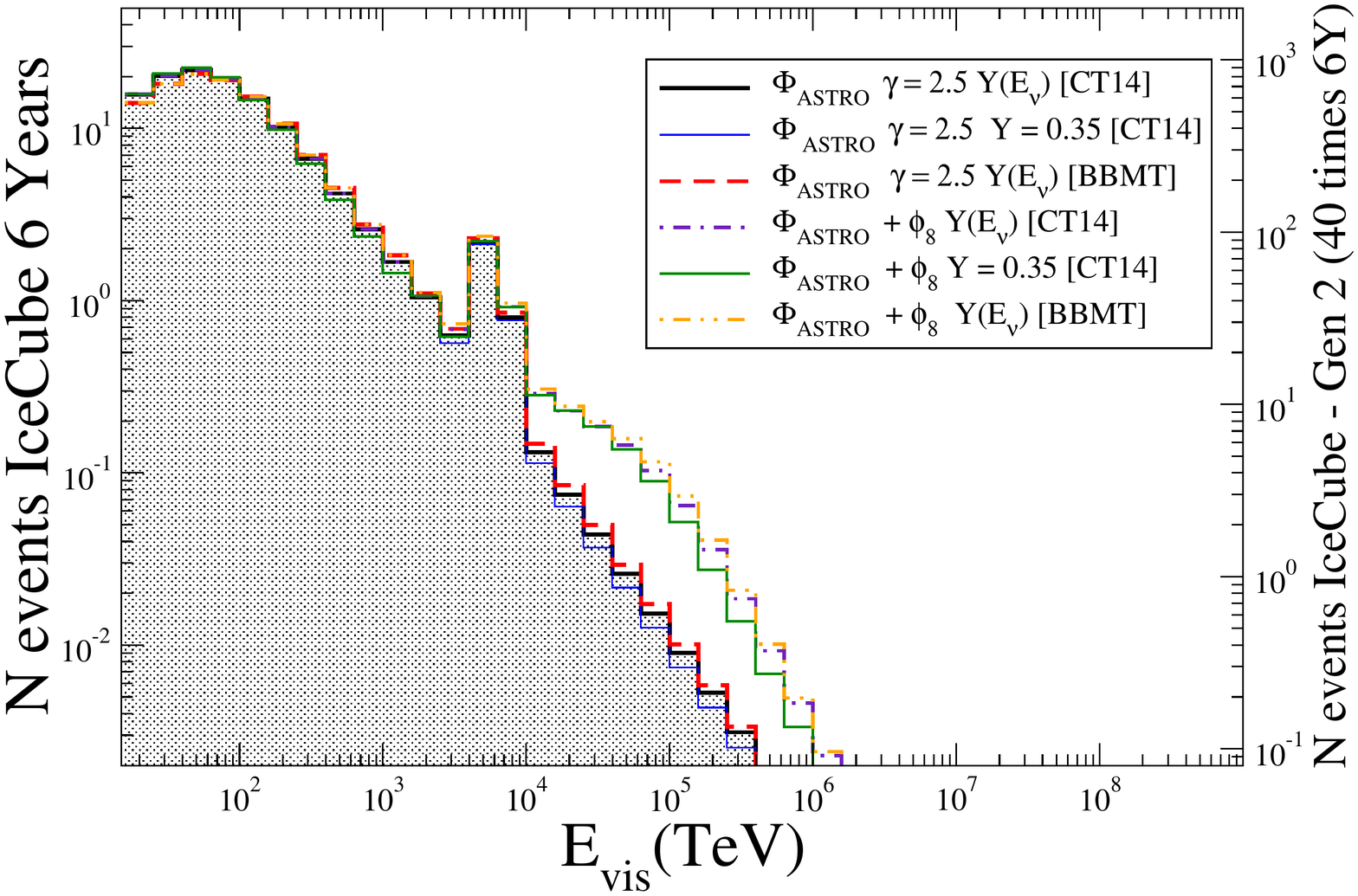} &
\includegraphics[scale=0.31]{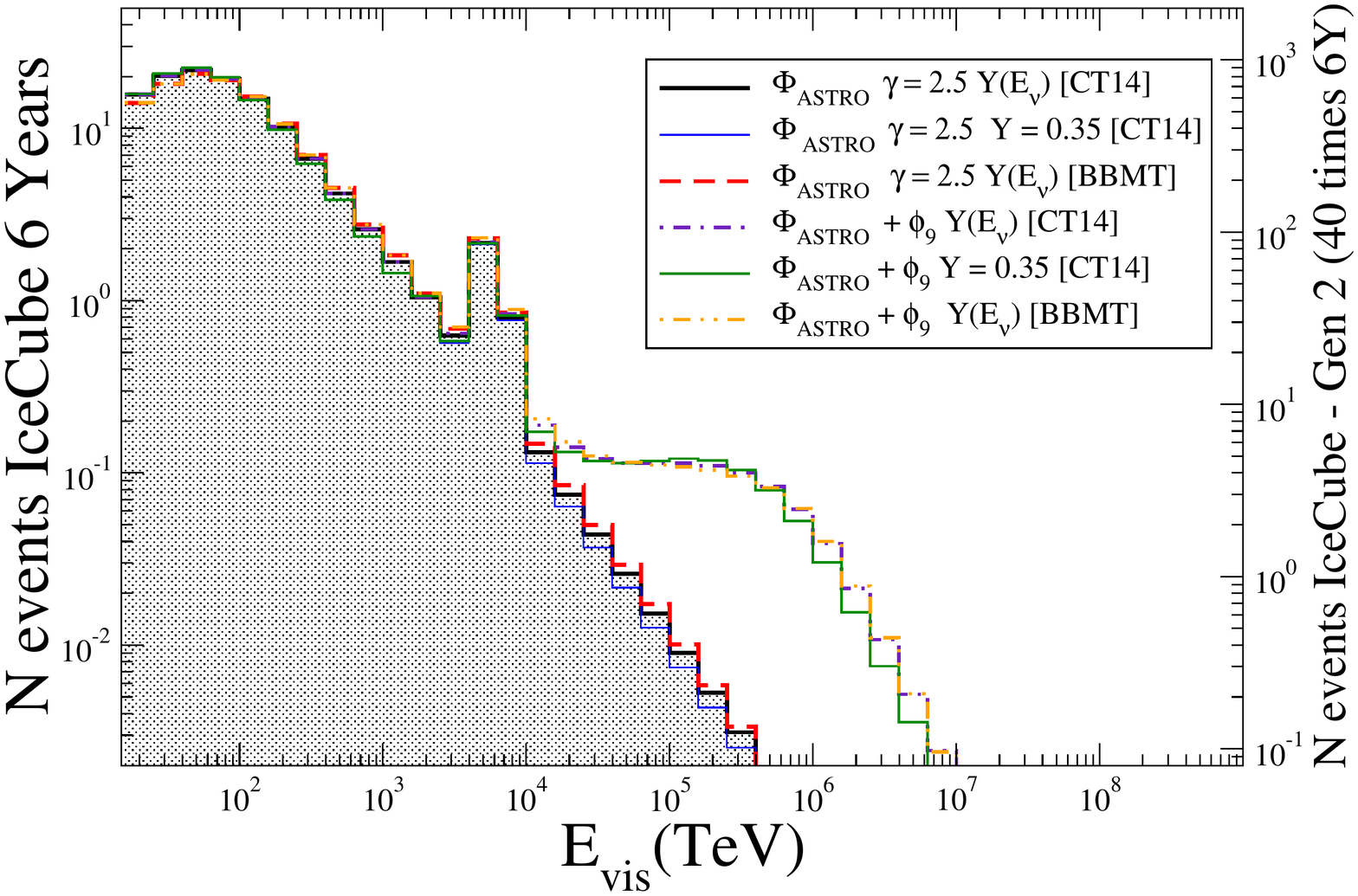} \\
(c) & (d)\\
\includegraphics[scale=0.31]{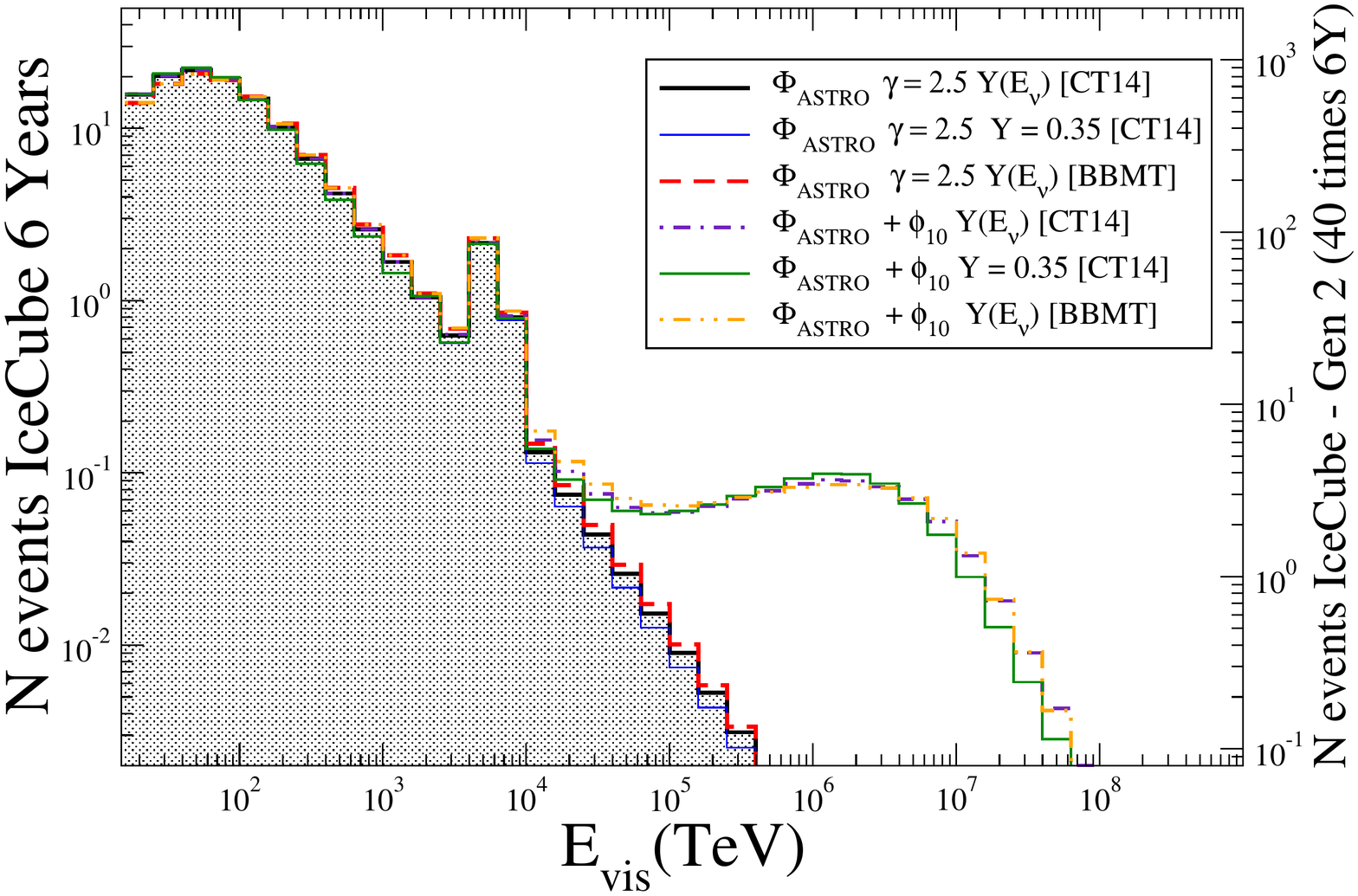} &
\includegraphics[scale=0.31]{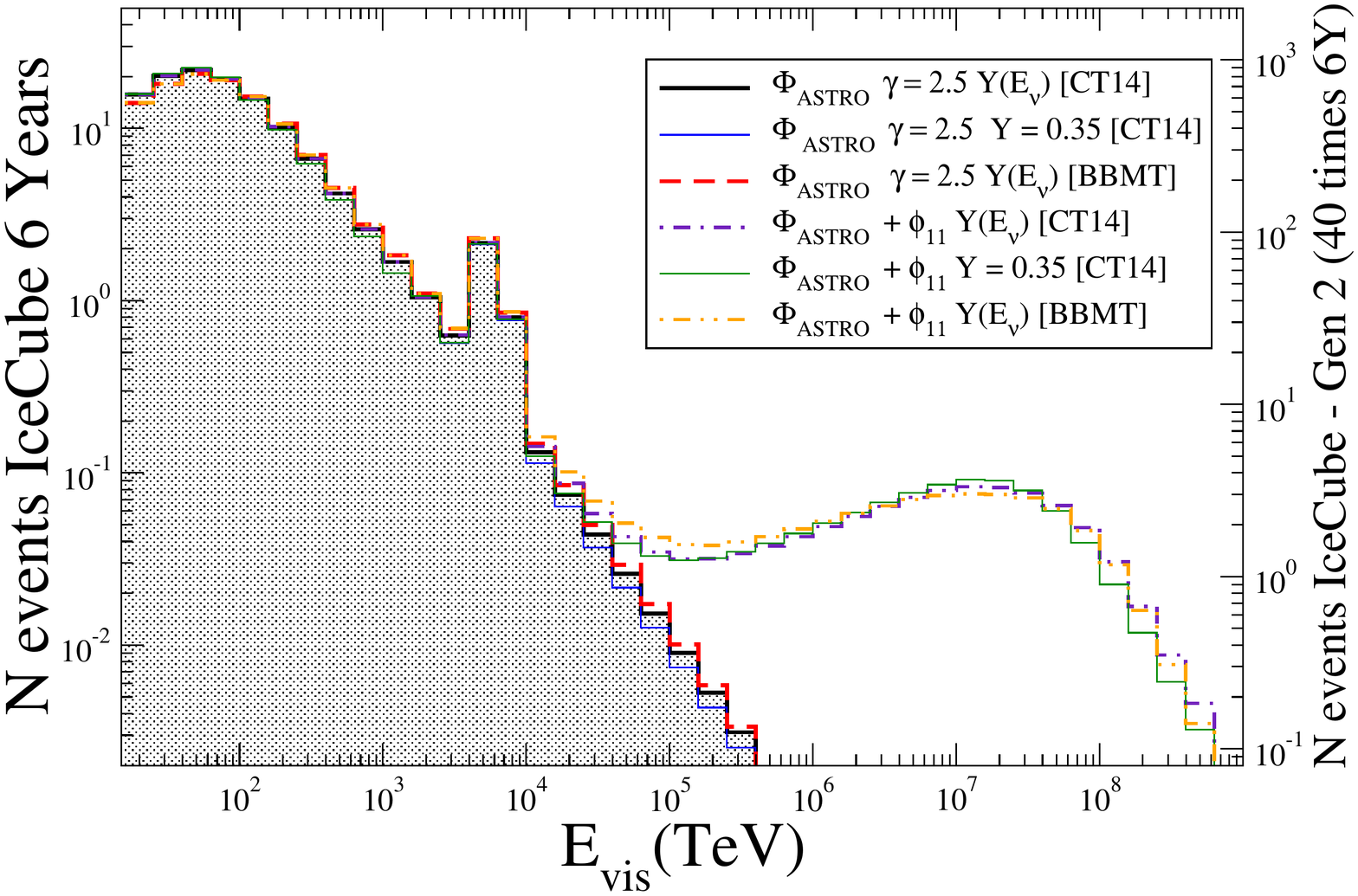} \\
(e) & (f)
\end{tabular}
\caption{Number of events in the IceCube detector as function of the visible energy for the distinct QCD models and fluxes considered. The left vertical axis refers to  6 years  the detector exposure, while in the right vertical axis represent the number of expected neutrino events is rescaled to the exposure of the  IceCube-Gen2. The standard astrophysical neutrino flux is described by the parameters  $\Phi_{0}=2.0 f.u.$ and the values of $\gamma$ indicated in the plots. {  In all  cases, the  normalization for the Super-Glashow  fluxes,  
$\phi_{0j}$, were adjusted to generate $\approx 1$ neutrino event at the energies above the Glashow resonance.} }
\label{Fig:NeveSG}
\end{center}
\end{figure}

{ Finally, let us estimate the impact of the nonlinear effects on the determination of the energy distribution of the neutrino events assuming the existence of the (still hypothetical)  Super-Glashow neutrino fluxes.} As discussed in the previous Section,   the hypothesis that the astrophysical neutrino flux has a new  component peaked at neutrino energies higher than the energy characteristic of the Glashow resonance { enhances} the occurrence of neutrino events at very high energies.  In Fig. \ref{Fig:NeveSG} (a), we compare the energy distribution of events at the IceCube observatory considering only the standard astrophysical neutrino flux as given in Eq. (\ref{Eq:flux}).  We consider two distinct scenarios for the standard flux: {(a)} {A {\it harder } spectrum, characterized by $\gamma = 2.5$, which is inspired by the best fit value of this parameter presented in Ref. \cite{IceCube:2020acn} for the case of cascades; and {(b)} A {\it softer } spectrum, which assumes $\gamma = 2.3$, and  approaches the best fit value in the  analysis of the muon tracks quoted in the same reference. Concerning the detector exposure,  we show our results for both the six-year sample, and for the planned extension of the observatory, the IceCube-Gen2. We found that case {(b)} is more likely to produce neutrino events at the ultrahigh-energy limit, and, at the same time, implies twice   the number of events at the resonance region. In particular, the case (b) predicts the double of events at the resonance region and an increasing by a factor 4 in the number of events for larger energies, mainly located at $E_{vis} \lesssim 10^{5}$ TeV.   %When the IceCube-Gen2 exposure is considered, our results for de case {\it a} ({\it b }) point to $\approx 120~ (\approx 250)$ events at the resonance region and  $\approx 10~ (\approx 40)$ events  above it, mainly located at $E_{vis}\le \approx 10^{5}$ TeV. 
The results presented in Fig.  \ref{Fig:Evis} indicate that cascade events with these energies are generated by neutrinos with $E_{\nu} \approx 10^{9}$ GeV, where we expect a suppression by $\approx 20 \%$ in the average inelasticity due to the nonlinear effects (See Fig.  \ref{Fig:CS}). As a consequence, nonlinear QCD effects are expected to become non-negligible in the description of the IceCube-Gen2 data.}

{ In what follows, we will assume that  the standard astrophysical flux is described by the harder spectrum discussed above, which describes the current IceCube data, and we will analyze the impact of a Super - Glashow flux $\phi_j$, described by distinct values of $j$ and that peaks for  neutrino energies above $10$ PeV. In our analysis, we have adjusted the normalization of the { Super-Glashow} flux in order to generate $\approx 1$ neutrino event  with visible energy above the Glashow resonance when the six-year exposure is considered. Our results are presented in  { panels b - f } of Fig. \ref{Fig:NeveSG}.   For $j = 7$ [panel (b)], we have obtained that the  number of events at neutrino energies of the order of the Glashow resonance is in agreement with the number observed in the presented exposure of the IceCube detector. Such conclusion is valid for the three  inelasticity scenarios considered.}  Also, as one can see,  from panels (b) and (c) respectively,  for energies around PeV there still is some degree of superposition between the distribution of neutrino events due to the standard flux with the {Super-Glashow} fluxes $\phi_{7}$ and $\phi_{8}$. On the other hand, the fluxes $\phi_{9}$ , $\phi_{10}$,  and $\phi_{11}$ generate events at energies far above the Glashow resonance. In Table \ref{Tab:03} we present a more detailed comparison between the predictions for the number of events for two distinct values of $E_{vis}$. For $\phi_{7}$ and $\phi_{8}$, one has that the large modification comes from the approximation $\langle Y\rangle = 0.35$. In fact,  our  results point out that the approximation of constant inelasticity combined with the DGLAP(CT14) model for the neutrino interaction tends to {  decrease the number of events in $\approx 10\% - 20\% $ in the ultrahigh-energy limit in comparison to the case where the inelasticity is assumed to be energy dependent.  On the other hand, the BBMT predictions for $\sigma_{\nu N}$ and $\langle Y (E_{\nu}) \rangle$ implies  a suppression in the number of events at both the peak and at the ultrahigh neutrino energy limit. This situation is seen  in the case of $\phi_{11}$ in Fig. \ref{Fig:NeveSG} (f). For $\phi_{9}$ and $\phi_{10}$ [See panels (d) and (e)], we found reductions of the order of $\approx 5\% - 10\% $ at the peak of the distribution.} These results indicate that the accurate description of the future  IceCube-Gen2 data cannot be performed assuming the approximation of $\langle Y(E_{\nu}) \rangle =  constant $. Moreover,  our results also indicate that if a  Super-Glashow flux is present,  the IceCube-Gen2 data could be sensitive to the nonlinear corrections on the neutrino-nucleon CC cross-section.

\begin{table*}[t] 
%\small
%\centering   
\begin{tabular}{|c|c|c|c|c|c|c|c|c|}
\hline
\multicolumn{3}{|l|}{\large \hspace{3cm} \bf $\Phi_{astro} + \phi_{7} $ } & \bf $E^{peak}_{vis} = 6.3 \times 10^{6}$ GeV  \\
\hline
\hline
&CT14 : $Y(E_{\nu})$& CT14 : $Y=0.35$& BBMT : $Y(E_{\nu})$  \\
\hline
N($E_{vis} = 10^{8}$ GeV)&$1.7\times 10^{-2}$ &$1.3\times 10^{-2}$ &$1.9\times 10^{-2}$ \\
\hline
N($E_{vis} = E^{peak}$)& 1.07 & 1.12& 1.14 \\
\hline
\hline
\hline
\multicolumn{3}{|l|}{\large \hspace{3cm}$\Phi_{astro} + \phi_{8} $ } & \bf $E^{peak}_{vis} = 1.6 \times 10^{7}$ GeV  \\
\hline
\hline
&CT14 : $Y(E_{\nu})$& CT14 : $Y=0.35$& BBMT : $Y(E_{\nu})$  \\
\hline
N($E_{vis} = 10^{8}$ GeV)&$6.4\times 10^{-2}$ &$5.2\times 10^{-2}$ &$7.3\times 10^{-2}$ \\
\hline
N($E_{vis} = E^{peak}$)&0.92 &0.92  & 0.97\\
\hline
\hline
\hline
\multicolumn{3}{|l|}{\large \hspace{3cm}$\Phi_{astro} + \phi_{9} $ } & \bf $E^{peak}_{vis} =  1.0\times 10^{8}$ GeV  \\
\hline
\hline
&CT14 : $Y(E_{\nu})$& CT14 : $Y=0.35$& BBMT : $Y(E_{\nu})$  \\
\hline
N($E_{vis} = 10^{8}$ GeV)& 0.11 & 0.12  & 0.10 \\
\hline
N($E_{vis} = E^{peak}$)& 0.11 & 0.12  & 0.10 \\
\hline
\hline
\hline
\multicolumn{3}{|l|}{\large \hspace{3cm}$\Phi_{astro} + \phi_{10} $ } & \bf $E^{peak}_{vis} = 1.0 \times 10^{9}$ GeV  \\
\hline
\hline
&CT14 : $Y(E_{\nu})$& CT14 : $Y=0.35$& BBMT : $Y(E_{\nu})$  \\
\hline
N($E_{vis} = 10^{8}$ GeV)&$5.9\times 10^{-2}$ &$6.0\times 10^{-2}$ &$ 6.4 \times 10^{-2}$ \\
\hline
N($E_{vis} = E^{peak}$)& $9.1 \times 10^{-2}$& $9.9 \times 10^{-2}$  & $8.5 \times 10^{-2}$ \\
\hline
\hline
\hline
\multicolumn{3}{|l|}{\large \hspace{3cm}$\Phi_{astro} + \phi_{11} $ } & \bf $E^{peak}_{vis} = 1.0 \times 10^{10}$ GeV  \\
\hline
\hline
&CT14 : $Y(E_{\nu})$& CT14 : $Y=0.35$& BBMT : $Y(E_{\nu})$  \\
\hline
N($E_{vis} = 10^{8}$ GeV)&$3.1\times 10^{-2}$ &$3.1\times 10^{-2}$ &$  3.8\times 10^{-2 }$ \\
\hline
N($E_{vis} = E^{peak}$)& $ 8.3\times 10^{-2}$&  $9.1\times 10^{-2}$ & $ 7.5\times 10^{-2}$ \\
\hline
\end{tabular}
\caption{The number of events in the IceCube detector for  6 years of the detector exposure, for the models and fluxes consider. {  In all  cases, the  normalization for the Super-Glashow  fluxes,   
$\phi_{0j}$, were adjusted to generate $\approx 1$ neutrino event at the energies above the Glashow resonance.}}
\label{Tab:03}
\end{table*}

\section{Summary}
\label{sec:conc}
One of the main goals of the IceCube observatory is the study of  UHE neutrino events, which are expected  to improve our understanding about the  origin, propagation, and interaction of  neutrinos. In recent years, several studies have focused on the use of the IceCube data as a way to constrain the energy behavior of the  astrophysical neutrino flux and  the neutrino - hadron cross section, which determine  the flux and event rate at the detector. 
Our main goal in this paper was to contribute to this current effort, by analysing the impact of the nonlinear effects on the average inelasticity and the energy and angular dependencies of the probability of absorption, which are important ingredients on the description of the number of events observed at the IceCube. In our analysis, we have assumed the BBMT model to treat these effects, which imply a {lower} bound for the magnitude of the neutrino - nucleon cross section.  Our results indicated that nonlinear effects, as estimated by the BBMT model, strongly reduce the average inelasticity with respect to the linear predictions and that  the energy dependence of $Y(E_{\nu})$ must be taken into the account in the accurate description of the future neutrino telescopes. Moreover, these results also indicated that the  determination of the incoming neutrino { energy} from the hadronic (leptonic) cascade at the limit of ultrahigh neutrino energy is sensitive to the description of the QCD dynamics at high energies. Regarding to the probability of absorption, $P_{shad}(E_{\nu},\theta_{z})$, we have demonstrated that the nonlinear effects are non-negligible at high energies when the amount of matter crossed by the neutrino is small ($\cos \theta_z \approx 0$). In addition, we also have investigated the impact of these effects on the determination of the normalization and spectral index using the HESE data. Our results indicated that  the description of the number of neutrino events in the   HESE  data  is sensitive to the description of the QCD dynamics. Finally, we also have considered the possibility that a (still hypothetical) super-Glashow flux is present at very high neutrino energies and estimated the number of events in the IceCube and IceCube - Gen2 detectors as function of the visible energy for the distinct QCD models and different assumptions for the calculation of the average inelasticity. Such analysis has pointed out that the
IceCube-Gen2 data can be sensitive to the nonlinear corrections on the neutrino-nucleon CC cross-section.
Therefore, the results obtained in this paper indicate that if the magnitude of the nonlinear effects at high energies is large, as predicted by the BBMT model, the description of the events in  future generations of neutrino observatories could be sensitive to these effects. Our analysis strongly motivate a more detailed study, using e.g.  the approaches recently discussed {   in Refs. \cite{Valera:2022ylt,Esteban:2022uuw,Garcia:2020jwr,{IceCube:2020acn}}}, which we plan to perform in a forthcoming paper.

\begin{acknowledgments}
This work was partially supported by INCT-FNA (Process No. 464898/2014-5).  V.P.G. was partially supported by the CAS President's International Fellowship Initiative (Grant No.  2021VMA0019) and by CNPq, CAPES  and FAPERGS. 
\end{acknowledgments}

%%%%%%%%%%%%%%%%%%%%%%

%\begin{widetext}

%\end{widetext}  
\end{document}